.
\documentclass[%
 reprint,
 superscriptaddress,
 nofootinbib,
 amsmath,amssymb,
 prc,
 floatfix
]{revtex4-2}

\usepackage{graphicx}
\usepackage{dcolumn}
\usepackage{bm}
\usepackage{xcolor}
\usepackage{wasysym}
\usepackage{physics}
\usepackage{amsmath}
\usepackage{subfigure}
\usepackage[pdftex,colorlinks,linkcolor = blue,urlcolor  = blue,citecolor = blue]{hyperref}

\newcommand{\be}{\begin{equation}}
\newcommand{\ee}{\end{equation}}
\newcommand{\bea}{\begin{eqnarray}}
\newcommand{\eea}{\end{eqnarray}}

\begin{document}

\preprint{APS/123-QED}

\title{Systematic study of the low-lying electric dipole strength in Sn isotopes and its astrophysical implications}

\author{M.~Markova}
\email{maria.markova@fys.uio.no}
\affiliation{Department of Physics, University of Oslo, N-0316 Oslo, Norway}

\author{A.~C.~Larsen}
\email{a.c.larsen@fys.uio.no}
\affiliation{Department of Physics, University of Oslo, N-0316 Oslo, Norway}

\author{P.~von Neumann-Cosel}%
\affiliation{%
 Institut f\"{u}r Kernphysik, Technische Universit\"{a}t Darmstadt, D-64289 Darmstadt, Germany
}%

\author{E.~Litvinova}%
\affiliation{%
 Department of Physics, Western Michigan University, Kalamazoo, Michigan 49008, USA
}%
\affiliation{%
National Superconducting Cyclotron Laboratory, Michigan State University, East Lansing, Michigan 48824, USA
}%
\affiliation{%
GANIL, CEA/DRF-CNRS/IN2P3, F-14076 Caen, France
}%

\author{A.~Choplin}
\affiliation{Institut d’Astronomie et d'Astrophysique, Université Libre de Bruxelles, CP 226, B-1050 Brussels, Belgium}

\author{S.~Goriely}
\affiliation{Institut d’Astronomie et d'Astrophysique, Université Libre de Bruxelles, CP 226, B-1050 Brussels, Belgium}

\author{S.~Martinet}
\affiliation{Institut d’Astronomie et d'Astrophysique, Université Libre de Bruxelles, CP 226, B-1050 Brussels, Belgium}

\author{L.~Siess}
\affiliation{Institut d’Astronomie et d'Astrophysique, Université Libre de Bruxelles, CP 226, B-1050 Brussels, Belgium}

\author{M.~Guttormsen}
\affiliation{Department of Physics, University of Oslo, N-0316 Oslo, Norway}

\author{F.~Pogliano}
\affiliation{Department of Physics, University of Oslo, N-0316 Oslo, Norway}

\author{S.~Siem}
\affiliation{Department of Physics, University of Oslo, N-0316 Oslo, Norway}

\date{\today}

\begin{abstract}
The $\gamma$-ray strength functions (GSF) and nuclear level densities (NLD) below the neutron threshold have been extracted for $^{111-113,116-122,124}$Sn from particle-$\gamma$ coincidence data with the Oslo method. The evolution of bulk properties of the low-lying electric dipole response has been investigated on the basis of the Oslo GSF data and results of a recent systematic study of electric and magnetic dipole strengths in even-even Sn isotopes with relativistic Coulomb excitation. The obtained GSFs reveal a resonance-like peak on top of the tail of the isovector giant dipole resonance, centered at $\approx$8 MeV and exhausting $\approx$2\% of the classical Thomas-Reiche-Kuhn (TRK) sum. For mass numbers $\geq$118 the data suggest also a second peak centered at $\approx$6.5 MeV. It corresponds to 0.1-0.5\% of the TRK sum rule and shows an approximate linear increase with the mass number. In contrast to predictions of the relativistic quasiparticle random-phase and time-blocking approximation calculations (RQRPA and RQTBA), no monotonous increase in the total low-lying $E1$ strength was observed in the experimental data from $^{111}$Sn to $^{124}$Sn, demonstrating rather similar strength distributions in these nuclei. 
The Oslo GSFs and NLDs were further used as inputs to constrain the cross sections and Maxwellian-averaged cross sections of $(n,\gamma)$ reactions in the Sn isotopic chain using TALYS. The obtained results agree well with other available experimental data and the recommended values from the JINA REACLIB, BRUSLIB, and KADoNiS libraries. Despite relatively small exhausted fractions of the TRK sum rule, the low-lying electric dipole strength makes a noticeable impact on the radiative neutron-capture cross sections in stable Sn isotopes. Moreover, the experimental Oslo inputs for the $^{121,123}$Sn$(n,\gamma)$$^{122,124}$Sn reactions were found to affect the production of Sb in the astrophysical $i$-process, providing new constraints on the uncertainties of the resulting chemical abundances from multi-zone low-metallicity Asymptotic Giant Branch  stellar models.

\end{abstract}

\maketitle
\section{\label{sec 1: introduction}Introduction}

 The study of the multipole electromagnetic response of atomic nuclei has always been an ultimate testing ground for unraveling a plethora of complex collective and single-particle excitation modes, their interplay, and driving physical mechanisms of nuclear interaction.
 Historically, one of the most well-studied modes of collective motion is the isovector giant dipole resonance (IVGDR), and the experimental and theoretical systematics on the IVGDR and its bulk properties are currently available for a wide range of mass numbers \cite{Eramzhyan1986,Kusnezov1999,Kleinig2008}.
 Within a macroscopic picture, this prominent feature is interpreted as a signature of out-of-phase dipole oscillations of all protons against all neutrons in the nucleus \cite{Migdal1944}. 
 
 In contrast to the IVGDR located at $10-20$ MeV in heavy nuclei, 
 the concentration of a weaker electric dipole strength in the vicinity of the neutron threshold, often referred to as the pygmy dipole resonance (PDR), is far less understood and keeps posing new questions regarding its origin and properties \cite{Paar2007,Savran2013,Bracco2019, Lanza2022}. A macroscopic interpretation of the PDR emerging from oscillations of excess neutrons, or a neutron skin, versus an isospin-saturated core \cite{Mohan1971} has been frequently adopted in publications since the 1970s, shifting the main focus of the experimental efforts in the past few decades to heavier, more neutron rich nuclei to test this interpretation~\cite{Adrich2005,Wieland2009, Wieland2018}. 
 However, this collective surface-motion picture and the degree of collectivity of involved transitions have been a matter of intense debates \cite{Paar2007, Litvinova2009, Roca-Maza2012, Reinhard2013, Bracco2019, Lanza2022}. Some studies suggest the physical mechanism behind the PDR to be a toroidal electric dipole mode instead of a neutron-skin oscillation\cite{Papakonstantinou2011,Repko2013,Derya2014,Neumann-Cosel2023}. 
 
 Another intensively discussed matter related to the PDR energy region is the isovector and/or isoscalar nature of observed structures \cite{Roca-Maza2012, Vretenar2012}.  
 The isospin splitting of the low-lying electric dipole response (LEDR) was, indeed, experimentally confirmed in complementary studies with  isoscalar and isovector probes in $^{124}$Sn, $^{138}$Ba, and $^{140}$Ce \cite{Endres2009, Endres2012, Pellegri2014, Krzysiek2016, Savran2018}. 
 Combined with self-consistent relativistic quasiparticle time-blocking approximation (RQTBA) and quasiparticle-phonon model (QPM) calculations, these experiments point towards the presence of two groups of transitions below the neutron threshold \cite{Endres2012, Savran2018}.
 The lower-lying group of states reveals a signature of a strong neutron contribution on the surface, whereas the higher-lying states are of more isovector nature, corresponding rather to the tail of the IVGDR. Furthermore, recent experiments on $^{208}$Pb \cite{Spieker2020} and $^{120}$Sn \cite{Weinert2021} with a deuteron probe combined with a QPM analysis provided insight into the one-particle-one-hole structure of the LEDR in these nuclei, revealing a similar structural splitting based on the contributing particle-hole configurations. 
 In general, an extensive investigation of the PDR region with complementary isoscalar and isovector probes in various inelastic scattering reactions as well as single-nucleon transfer reactions is pivotal to break down the complex LEDR structure of nuclei within different mass regions.

 The tin isotopic chain is probably one of the best studied cases both experimentally and theoretically (see, e.g.,  the review articles in ~\cite{Savran2013,Lanza2022}). 
 For the tin isotopes, the LEDR is available from  nuclear resonance fluorescence (NRF) studies \cite{Govaert98,Ozel-Tashenov2014, Muscher2020}, experiments with $\alpha$~\cite{Endres2012}, $^{17}$O~\cite{Pellegri2014}, and deuteron~\cite{Weinert2021} probes, Coulomb dissociation \cite{Adrich2005, Klimkiewicz2007} and Coulomb excitation \cite{Krumbholz2015,Bassauer2020b} experiments. A first attempt to extract the systematics of the PDR observed in the GSFs of $^{116-119,121,122}$Sn measured with the Oslo method was presented in Ref.~\cite{Toft2011}. 
 With new experimental information available on $^{120,124}$Sn \cite{Markova2021, Markova2022} and $^{111-113}$Sn \cite{Markova2023}, a combination of the Oslo data and the recently published strengths for even-even Sn isotopes from Coulomb excitation experiments \cite{Bassauer2020b} permits a consistent extensive study on the evolution of the LEDR, covering eleven Sn isotopes from $^{111}$Sn to $^{124}$Sn.  

The connection of the PDR strength to the neutron-skin thickness in neutron-rich nuclei, suggested by the neutron oscillation picture,
triggered attempts to provide experimental constraints on
the symmetry-energy term in the equation of state \cite{Piekarewicz2006, Klimkiewicz2007,Piekarewicz2011,Roca-Maza2011}, with implications  for the characteristics of neutron stars \cite{Horowitz2001,Fattoyev2012}. 
While this connection is under debate \cite{Reinhard2013,Reinhard2014}, influence of the enhanced $E1$ strength close to neutron threshold on the astrophysical radiative neutron-capture rates is less ambiguous \cite{Litvinova2009, Goriely2004}.
An increased probability of radiative neutron capture due to the boosted GSF within the PDR region might impact the element production in the rapid neutron-capture process ($r$ process), responsible for creating $\sim 50\%$ of elements heavier than Fe in the universe. 
However, assessing the importance of the PDR in the $r$-process nucleosynthesis is difficult due to a lack of experimental constraints for very neutron-rich nuclei and a large spread in theoretical predictions of the PDR strength.
To provide radiative neutron-capture rates for $r$-process reaction network calculations, the statistical Hauser–Feshbach model is employed \cite{Hauser1952, Rauscher2000, Koning2023}. 
This model calls for consistently extracted experimental data on the nuclear level densities (NLD) and $\gamma$-ray strength functions (GSF), or average reduced $\gamma$-transition probability, for experimentally accessible cases to constrain the available theoretical models.

In addition to investigating the impact of the PDR on the $r$ process,
dipole strength distributions below the neutron threshold in stable isotopes are of general interest 
for the slow ($s$) neutron-capture process. 
The majority of the stable Sn isotopes originate predominantly from the $s$ process \cite{Macklin1962,Wisshak1996,Koehler2001}, with $^{121}$Sn and $^{123}$Sn being potential candidates for the $s$-process branching point nuclei (see Refs.~\cite{Kaeppeler1993,Diehl2018}, respectively). 
Moreover, isotopes heavier than $^{120}$Sn might be involved in the main flow of the intermediate ($i$) neutron-capture process, as discussed by Goriely \textit{et al.} \cite{Goriely2021}. 
The Oslo method 
enables the extraction of both key nuclear inputs, the NLDs and GSFs, for statistical calculations within the Hauser–Feshbach framework. 
Therefore, the method provides experimental constraints on the radiative neutron-capture reaction rates of interest for all the three above-mentioned nucleosynthesis processes.

The paper is outlined as follows. Section~\ref{sec 2: Method and Experiment} describes the details of experiments on the Sn isotopes performed at the Oslo Cyclotron Laboratory (\ref{subsec 2.1: Experiments}) and the Oslo method (\ref{subsec 2.2: Oslo method}). 
In Sec.~\ref{sec 3: Results}, the extracted NLDs (\ref{subsec 3.1: NLDs}), GSFs, and the systematics of the bulk properties of the low-lying $E1$ strength (\ref{subsec 3.2: GSFs and PDRs}) are presented. 
Sec.~\ref{sec 4: Exp plus theory} focuses on the comparison of this systematics with model predictions. 
The neutron-capture cross sections and Maxwellian-averaged cross sections are presented and discussed together with the potential role of the LEDR in Sec.~\ref{sec 5: NCCS and MACS}. 
In Sec.~\ref{sec 6: i process},  $i$-process calculations in Asymptotic Giant Branch (AGB) stars are presented, and the impact of the new experimentally constrained rates on the production of the elements in the Sn region is discussed. 
Finally, the main findings of this work are summarized in Sec.~\ref{sec 6: Conclusion}.
                                         
\section{\label{sec 2: Method and Experiment}Data and methodology}
\subsection{\label{subsec 2.1: Experiments} Setup and experimental details}
Eleven tin isotopes, $^{111-113,116-122,124}$Sn, were studied in nine experimental campaigns taking place at the Oslo Cyclotron Laboratory (OCL) in the period from 2003 to 2022. All nuclei were studied in light-particle-induced reactions with $p$, $d$, and $^{3}$He beams delivered by the MC-35 Scanditronix cyclotron with the main objective of extracting particle-$\gamma$ coincidence events for a further analysis with the Oslo method. In all cases, the configuration of the setup involved a scintillator $\gamma$-ray detector array surrounding the target chamber and a Si particle-telescope system placed either in forward or backward angles with respect to the beam direction. 

\begin{table*}[t]
\caption{\label{tab:table_1}Characteristics of the experiments on Sn isotopes performed at the OCL. The given angles refer to the particle scattering angles with respect to the beam direction. }
\begin{ruledtabular}
\begin{tabular}{lccccllll}
Target & Thickness & Enrichment & Reaction & Beam energy & Beam intensity & Angles & Year & Setup  \\ 
 & (mg/cm$^2$) & (\%) &  & (MeV) & (nA) & ($^{\circ}$) & & \\
\noalign{\smallskip}\hline\noalign{\smallskip}
$^{112}$Sn & 4.0  & 99.8 & ($p,d\gamma$) & 25.0 & $\approx 1.0–1.5$ & 126-140 & 2013 & SiRi+CACTUS\\
           &      &      & ($p,p^{\prime}\gamma$) & 16.0 & $\approx 1.0–1.5$ & 126-140 & 2013 & SiRi+CACTUS\\ 
           &      &      & ($d,p\gamma$) & 11.5 & $\approx 0.5–0.7$ & 126-140 & 2014 & SiRi+CACTUS\\
\noalign{\smallskip}\hline\noalign{\smallskip}
$^{117}$Sn & 2.1 & 92.0 & ($^{3}$He,$\alpha\gamma$) & 38 & $\approx 1.5$ & $\approx 42.5-47.5$ & 2003 & Stand. Si + CACTUS \\
 &  &  & ($^{3}$He,$^{3}$He$\gamma$)\footnotemark[1] & 38 & $\approx 1.5$ & $\approx 42.5-47.5$ & 2003 & Stand. Si + CACTUS \\
 &  &  & ($p,p^{\prime}\gamma$)\footnotemark[2] & 16 & $\approx 2.8$ & 126-140 & 2019 & SiRi + OSCAR \\
\noalign{\smallskip}\hline\noalign{\smallskip}
$^{119}$Sn &  1.6 & 93.2 & ($^{3}$He,$\alpha\gamma$) & 38 & $\approx 1.5$ & $\approx 42.5-47.5$ & 2008 & Stand. Si + CACTUS \\
 &   &  &  ($^{3}$He,$^{3}$He$\gamma$)\footnotemark[1] & 38 & $\approx 1.5$ & $\approx 42.5-47.5$ & 2008 & Stand. Si + CACTUS \\
 &   &  &  ($p,p^{\prime}\gamma$)\footnotemark[2] & 16 & $\approx 0.6-0.8$ & 126-140 & 2022 & SiRi + OSCAR \\
\noalign{\smallskip}\hline\noalign{\smallskip}
$^{120}$Sn & 2.0  & 99.6 & ($p,p^{\prime}\gamma$) & 16 & $\approx 3.0-4.0$ & 126-140 & 2019 & SiRi + OSCAR \\
\noalign{\smallskip}\hline\noalign{\smallskip}
$^{122}$Sn & 1.43  & 94 & ($^{3}$He,$\alpha\gamma$) & 38 & $\approx 0.2$ & 40-54 & 2010 & SiRi+CACTUS \\
 &   &  & ($^{3}$He,$^{3}$He$\gamma$) & 38 & $\approx 0.2$ & 40-54 & 2010 & SiRi+CACTUS \\
\noalign{\smallskip}\hline\noalign{\smallskip}
$^{124}$Sn & 0.47  & 95.3 & ($p,p^{\prime}\gamma$) & 16 & $\approx 3.0-4.0$ & 126-140 & 2019 & SiRi + OSCAR \\
\end{tabular}
\end{ruledtabular}
\footnotetext[1]{Not used in the present work.}
\footnotetext[2]{Remeasured.}
\end{table*}

The first experiments on $^{117}$Sn and $^{119}$Sn, aiming at studying $^{116,117}$Sn and $^{118,119}$Sn, respectively, utilized the ($^{3}$He,$\alpha \gamma$) and ($^{3}$He,$^{3}$He$^\prime\gamma$) reactions. 
These experiments were performed with 38-MeV $^{3}$He beams and exploited eight standard $\Delta E-E$ Si telescopes with $\approx$140-$\mu$m-thick $\Delta E$  and $1500$-$\mu$m-thick $E$ counters. The telescopes were placed within the target chamber at 45$^{\circ}$ with respect to the beam direction, as a compromise between reducing the contribution from elastic scattering and still having sufficiently large cross sections for the reactions of interest. 
Collimators were placed in front of the Si detectors to reduce the uncertainty in the scattering angle. 
This collimation led to a significantly reduced solid-angle coverage of $\approx$0.72\% of 4$\pi$ and $\approx$1.5\% of 4$\pi$ for the collimators with circular ($^{117}$Sn) and squared openings ($^{119}$Sn), respectively.

To improve the solid-angle coverage while maintaining a reasonable angular resolution, a custom-designed Si telescope ring (SiRi) was installed in 2011 \cite{Guttormsen2011}. 
The SiRi system was used in the experimental campaigns to study $^{120-122,124}$Sn as well as remeasuring $^{117,119}$Sn in $2019-2022$ in the ($p,p^{\prime}\gamma$) and $^{121,122}$Sn in ($^{3}$He,$\alpha \gamma$) and ($^{3}$He,$^{3}$He$^\prime \gamma$) reactions. 
SiRi is comprised of eight trapezoidal-shaped $\Delta E$-$E$ modules with 130-$\mu$m-thick $\Delta E$ layers and 1550-$\mu$m-thick $E$ layers. 
The former are additionally segmented into eight curved pads. The coverage of scattering angles in SiRi is either $40-54^{\circ}$ in the forward or $126-140^{\circ}$ in the backward position of the detector array, with a 2$^{\circ}$ azimuth angle window per each pad. With SiRi, the solid-angle coverage increased approximately 10 times as compared to the previous telescope system, while keeping a sufficient energy resolution.

Both the older Si detector system and SiRi make use of the $\Delta E$-$E$ technique to differentiate between the observed reaction channels. 
The typical energy resolution for the experiments with the 38-MeV $^{3}$He beam performed with the older setup was $\approx 250-350$ keV Full Width at Half Maximum (FWHM), determined from the fit to the elastic peaks in the ($^{3}$He,$^{3}$He$^{\prime}$) and ($^{3}$He,$\alpha$) channels. 
With SiRi, the  energy resolution is $\approx 150-200$ keV for the same experimental conditions, and $\approx 300$ keV for the 11.5-MeV deuteron and 20-MeV proton beams in the $(d,p)$ and $(p,d)$ channels, respectively (due to the large thickness of the $^{112}$Sn target). 
The best resolution of $\approx 100$ keV was achieved in the 
experiments using 16-MeV protons with SiRi. 
Besides the reaction channel, the beam energy, and intrinsic resolution of the $E-\Delta E$ modules, the excitation-energy resolution was also affected by the beam-energy resolution (the beam-spot size varied significantly in the 
experiments).

All the ($^{3}$He,$^{3}$He$^\prime\gamma$) and ($^{3}$He,$\alpha \gamma$) experiments on $^{117,119,122}$Sn as well as the ($p,p^{\prime}\gamma$), ($p,d\gamma$), and ($d,p\gamma$) experiments on $^{112}$Sn were performed with the detector array CACTUS \cite{Guttormsen1990_CACTUS}. CACTUS consisted of 28 spherically distributed cylindrical 5$^{\prime\prime}\times$ 5$^{\prime\prime}$ NaI(Tl) detectors, where each detector was additionally collimated with conical Pb collimators. 
Using a $^{60}$Co source, the total efficiency of CACTUS and its energy resolution at $E_{\gamma}=1332$ keV were estimated to be 15.2(1)\% and $\approx 6.8\%$, respectively. 

In 2019, the CACTUS detectors were replaced by OSCAR (Oslo SCintillator ARray),  a $\gamma$-ray detector array of 30 cylindrical large-volume 3.5$^{\prime\prime}\times$ 8$^{\prime\prime}$ LaBr$_{3}$(Ce) crystals \cite{Ingeberg2020, Zeiser2020}. 
OSCAR provides a significantly improved  energy resolution and excellent timing properties for selecting particle-$\gamma$ events. 
The total efficiency in the most recent 
experiments is $\approx 40\%$, with the energy resolution of $\approx 2.2\%$ at $E_{\gamma}=1332$ keV.

In the period between 2003 and 2022, $^{117}$Sn and $^{119}$Sn were measured twice; first, with the ($^{3}$He,$^{3}$He$^\prime \gamma$) reaction using the old setup configuration (standard Si telescopes + CACTUS), and later with SiRi and OSCAR using the ($p,p^{\prime}\gamma$) reactions.  
Due to a fairly good agreement between the new and the old data sets, and considering the improved energy resolution and timing with the new setup, we choose to show only the ($p,p^{\prime}\gamma$) data for $^{117}$Sn and $^{119}$Sn in the present work. 
The data processing before the application of the Oslo method for these two experiments is identical to the one described in detail for $^{120,124}$Sn in Ref.~\cite{Markova2022}. 
All relevant parameters for the above-mentioned experiments are outlined in Table \ref{tab:table_1}.

Using the known kinematics of the studied reactions, the energy deposited in the particle Si detectors was converted into the initial excitation energy $E_i$ of the residual nucleus. 
By applying proper gates on the outgoing particle species and the time spectra, 
the background-subtracted particle-$\gamma$ coincidence events were extracted. 
A more in-depth discussion of the experimental details, calibration, and event selection in each case is presented in Refs.~\cite{Markova2023} for $^{111-113}$Sn, Refs.~\cite{Agvaanluvsan09-1,Agvaanluvsan2009-2} for $^{116}$Sn, Ref.~\cite{Toft2010} for $^{118}$Sn, Ref.~\cite{Markova2022} for $^{120,124}$Sn,and Ref.~\cite{Toft2011} for $^{121,122}$Sn. 
Calibrations of the excitation and $\gamma$-ray energies for the $^{121,122}$Sn data sets were revised and improved compared to the earlier published results \cite{Toft2011}.

In the next step, the recorded $\gamma$-ray spectra are corrected using the detector response functions of either CACTUS or OSCAR to obtain the unfolded spectra. 
For all cases, the unfolding was done with the same iterative technique described in detail in Ref.~\cite{Guttormsen1996}. The method is based on a consecutive correction of the trial function for the unfolded spectra,  
until they match with the original raw spectra within the experimental uncertainties. 
To avoid introducing any strong artificial fluctuations while still preserving the original statistical fluctuations, the Compton subtraction technique was also applied in each case. 
The details of this procedure are presented in Refs.~\cite{Guttormsen1996, Larsen2011}.  

\begin{figure}[t]
\includegraphics[width=1.0\linewidth]{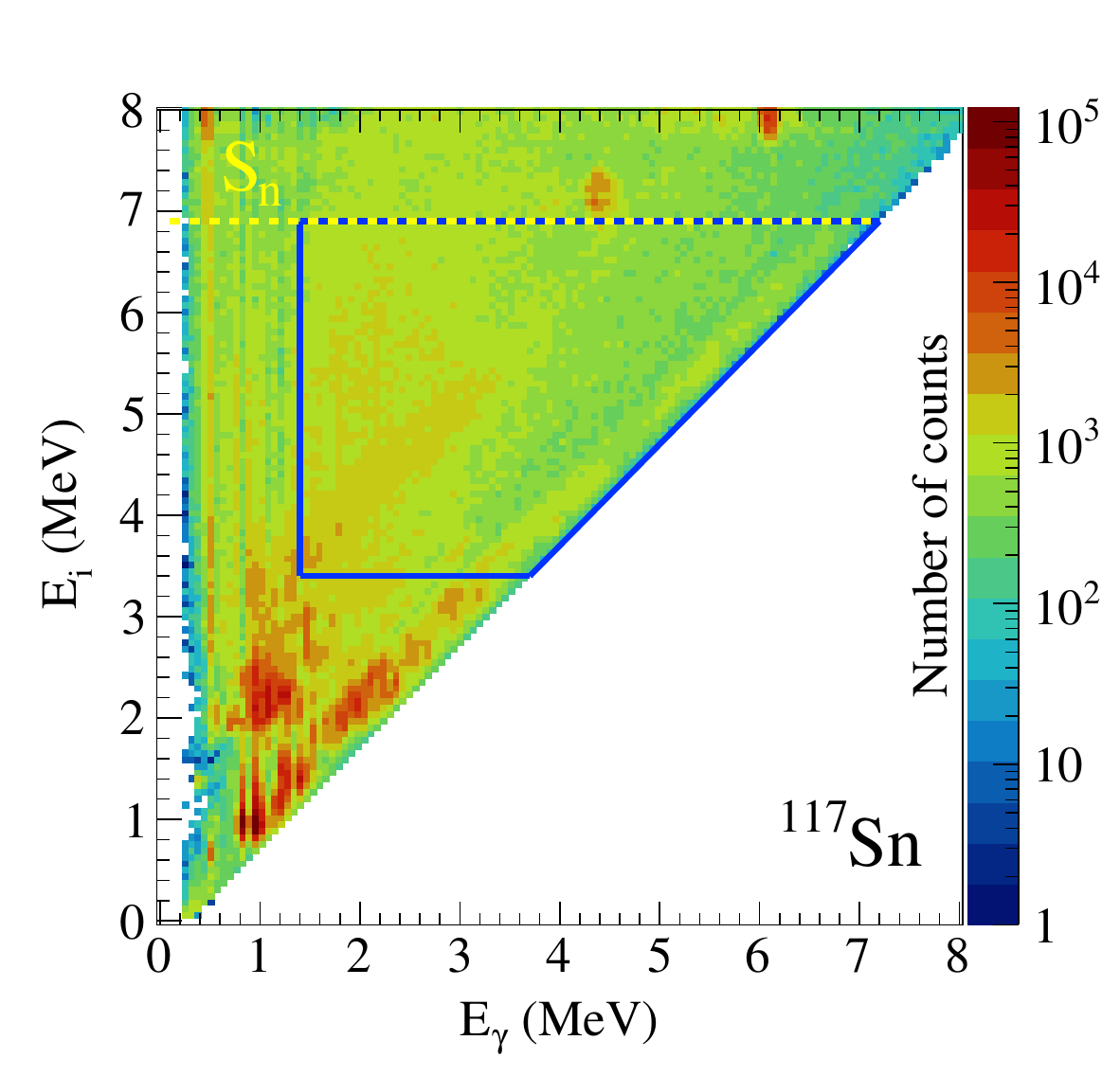}
\caption{\label{fig: LDs}
The first-generation matrix for $^{117}$Sn extracted in the ($p,p^{\prime}\gamma$) experiment. The yellow dashed line corresponds to the neutron separation energy, while the blue solid lines confine the area used for the Oslo method. The bin size is 64 keV$\times$64 keV.
}
\end{figure}

The last step prior to the extraction of the NLD and GSF is to determine the distribution of first-generation $\gamma$ rays for each excitation-energy bin, i.e., $\gamma$ rays stemming directly from a nucleus decaying with an initial excitation energy $E_i$. 
The distribution $P(E_{\gamma},E_i)$ of such first-generation -- or primary -- $\gamma$ rays, is proportional to the probability of $\gamma$ decay from initial excitation energies $E_i$ to the final levels $E_f=E_i-E_{\gamma}$, i.e., to the corresponding average branching ratios of the levels within the excitation-energy bin $E_i$. 
An important assumption for the extraction of the primary $\gamma$ rays is that the decay pattern of the excited levels within the $E_i$ bin is independent of the way they were populated (either directly through the reactions or via the decay of higher-lying states). 
Then, the $\gamma$ cascades at each initial excitation-energy bin $E_i$ are expected to contain the same transitions as those in the bins below, except for the primary $\gamma$ transitions. 
By use of the iterative first-generation method (see Ref.~\cite{Guttormsen1987} for details), the non-primary $\gamma$ transitions are successively removed from each initial excitation-energy bin below the neutron threshold. 
The above-mentioned assumption is expected to hold for relatively high excitation energies where the spin distribution is approximately equal for neighboring $E_i$ bins. 
This imposes a lower limit on the initial excitation energy for the further data analysis with the Oslo method, while an upper limit is set by the neutron separation energy $S_n$ in each case. 
In addition, an over-subtraction of $\gamma$ transitions at low $\gamma$ energies seen in all data sets limits $\gamma$-ray energy to $E_{\gamma}\gtrsim 1 -2$ MeV. 
A thorough discussion of the application of the first-generation method and its limitations is presented in  Ref.~\cite{Larsen2011}. 
For the most recent experiments on $^{117}$Sn and $^{119}$Sn, these limits were set to $3.4 \lesssim E_i \lesssim 6.9$ MeV, $E_{\gamma}\gtrsim 1.4$ MeV and $4.0 \lesssim E_i \lesssim 6.5$ MeV, $E_{\gamma}\gtrsim 1.8$ MeV, respectively. 
Discussions of the chosen $E_i$ and $E_{\gamma}$ limits for the subsequent Oslo method analysis of other Sn isotopes can be found in Refs.~\cite{Markova2023, Agvaanluvsan09-1,Agvaanluvsan2009-2, Toft2010,Markova2022,Toft2011}.

\subsection{\label{subsec 2.2: Oslo method}Extraction of NLDs and GSFs}

As mentioned in the previous section, the first-generation matrix reflects a distribution of decay probabilities from the levels within each  initial excitation-energy bin $E_i$ to the final levels $E_f$ via $\gamma$ transitions with energies $E_{\gamma}=E_i-E_f$. 
The Oslo method exploits this fact to perform a decomposition of the primary matrix $P(E_{\gamma},E_i)$ into the $\gamma$ transmission coefficient  $\mathcal{T}_{i\rightarrow f}$ and the density of final levels $\rho_f$:

\begin{equation}
\label{eq:1}
    P(E_{\gamma},E_i)\propto \mathcal{T}_{i\rightarrow f}\cdot\rho_f.
\end{equation}

This decomposition is supported by both Fermi’s golden rule and the Hauser-Feshbach theory of statistical reactions (see Ref.~\cite{MIDTBO_1} and Ref.~\cite{MIDTBO_2}, respectively, for  detailed derivations). 
Being valid only for the compound excited states, this relation is expected to hold well for the chosen excitation-energy ranges. 

The transmission coefficient in Eq.~(\ref{eq:1}) depends on both the initial and the final excitation energy, making it practically impossible 
to disentangle $\mathcal{T}_{i\rightarrow f}$ and $\rho_f$ in the factorized first-generation matrix.
For this reason, the generalized Brink-Axel hypothesis \cite{Brink1955, Axel1962} needs to be employed in the Oslo method to reduce the dependence of the transmission coefficient on the initial and final excitation energy 
to a dependence on $E_{\gamma}$ only ($\mathcal{T}_{i\rightarrow f}\rightarrow \mathcal{T}(E_{\gamma})$). 
In its most commonly used, generalized form, this hypothesis states that the GSF (and, thus, the $\gamma$ transmission coefficient proportional to it) is independent of excitation energy, spin, and parity of the initial and final levels. Even though the Brink-Axel hypothesis was originally formulated for the IVGDR region, it is commonly applied also in the PDR region at lower excitation energies, as it significantly simplifies any calculations involving photon absorption and emission \cite{Capote2009}. 
Even though the discussion around the applicability of the Brink-Axel hypothesis in this region involves cases where it holds well (e.g. \cite{Guttormsen2016, Martin2017}) as well as  cases where it seems questionable (e.g., Refs.~\cite{Isaak2013,Sieja2023}), 
it has been shown that 
for the Oslo method analysis of nuclei in different mass regions (including Sn isotopes) it 
is a reasonable assumption 
\cite{Larsen2011, Guttormsen2016, Campo2018, Markova2021, Markova2022}. 

To extract the NLD $\rho_f=\rho(E_i-E_{\gamma})$ and the $\gamma$ transmission coefficient $\mathcal{T}(E_{\gamma})$,  the first-generation matrix $P(E_{\gamma},E_i)$ 
is approximated in an iterative $\chi^2$-fit procedure with the following ``theoretical'' matrix \cite{Schiller2000}:
\begin{equation}
\label{eq:2}
P_{th}(E_{\gamma}, E_i)=\frac{\mathcal{T}(E_{\gamma})\rho(E_i-E_{\gamma})}{\sum_{E_{\gamma}=E_{\gamma}^{min}}^{E_i}\mathcal{T}(E_{\gamma})\rho(E_i-E_{\gamma})}.
\end{equation}
Prior to this step, the first-generation spectra are normalized to unity for each $E_i$ bin. 
The details of this procedure and the propagation of statistical, unfolding, and first-generation method uncertainties are outlined in Ref.~\cite{Schiller2000}. 
It has been repeatedly shown to converge well in each case \cite{Guttormsen2022, Larsen2023, Markova2023}, and the resulting  matrix $P_{th}(E_{\gamma}, E_i)$ reproduces the experimental spectrum quite well for each excitation-energy bin within the chosen limits. 

The obtained functions $\rho(E_i-E_{\gamma})$ and $\mathcal{T}(E_{\gamma})$ of $P_{th}(E_{\gamma}, E_i)$ provide the best fit of the experimental spectra and represent the solutions for the  experimental NLD and the $\gamma$ transmission coefficient. 
However, as shown in Ref.~\cite{Schiller2000}, although the variation of individual data points with respect to their neighboring points is uniquely determined by the fit, the solutions
can be modified with arbitrary chosen scaling parameters $A$ and $B$ and a slope parameter $\alpha$, providing an equally good fit to the experimental primary spectra through the following transformations:
\begin{equation}
\label{eq:3}
\begin{split}
    \tilde{\rho}(E_i-E_{\gamma})=&A\rho(E_i-E_{\gamma})\exp[\alpha (E_i-E_{\gamma})],\\
    \tilde{\mathcal{T}}(E_\gamma)=&B\mathcal{T}(E_\gamma)\exp[\alpha E_{\gamma}].
\end{split}
\end{equation}

To extract the ``true" physical NLD and the $\gamma$ transmission coefficient, the parameters $A$, $B$, and $\alpha$ must be constrained with external experimental data by following  normalization procedures as presented in the two subsequent sections. 

To ensure a fully consistent normalization procedure, all the Sn nuclei considered here were revisited and renormalized using the most updated experimental information and the same model approaches for the normalization. 
The main objective of this part of the analysis was not only to make use of the updated external data, but also to apply a consistent model approach for the spin distribution, supported by the most recent experimental and theoretical works \cite{Voinov2023,Hilaire2023}.  As was shown in the earlier publications, the latter yields a reasonable agreement of the Oslo method NLDs and GSFs with other experimental results \cite{Markova2021, Markova2023}. 
Also, a comparison with the inelastic relativistic proton scattering data [(p,p$^{\prime}$) for short] \cite{Bassauer2020b} providing the GSFs for the even-even $^{112,114,116,118,120,124}$Sn isotopes serves as a benchmark for the slope parameter $\alpha$ shared by the NLD and GSF as well as the absolute normalization of the strength.

\subsubsection{\label{subsubsec 2.2.1: NLD normalization}Normalization of the NLDs}

To determine the slope $\alpha$ and the absolute value $A$, the NLD solutions from Eq.~(\ref{eq:3}) are anchored to known low-lying excited states and  the NLD at the neutron separation energy $\rho(S_n)$.
We follow the same normalization procedure as presented in the latest publication on $^{111-113}$Sn \cite{Markova2023}. 
The most recent compilation of discrete states \cite{ensdf} was used for all isotopes. 
To estimate the total NLD at the neutron separation energy, $\rho(S_n)$, the average neutron-resonance spacing  $D_0$ ($s$-wave neutrons) or $D_1$ ($p$-wave neutrons) from neutron resonance experiments can be used. 
For seven out of eleven studied isotopes, namely $^{113,116-121}$Sn, such data on neutron resonances are available \cite{Mughabghab18}. 
For $s$-wave resonances, and for a target with non-zero target spin, the $D_0$ value can be written as:
\begin{equation}
\label{eq:4}
    \frac{1}{D_0}=\frac{1}{2}\bigl(\rho(S_n, J_t+1/2)+\rho(S_n, J_t-1/2)\bigr),
\end{equation}
where $J_t$ is the ground-state spin of the target (sample) in the neutron resonance experiments. 
The factor $1/2$ is due to the assumption that levels with positive and negative parities contribute equally to the NLD in the vicinity of $S_n$ \cite{Schiller2000,Larsen2011}. 
This assumption was shown to hold well for this excitation-energy region \cite{Toft2010,Larsen2011}. 
To calculate the \textit{total} NLD $\rho(S_n)$, we exploit that the \textit{partial} NLD for a given spin $J$ can be found through the relation $\rho(E_x, J) = g(E_x, J)\rho(E_x)$, where $g(E_x, J)$ is the spin distribution from Refs.~\cite{Ericson58,Gilbert65}:
\begin{equation}
\label{eq:5}
 g(E_x, J) \simeq \frac{2J+1}{2\sigma^2(E_x)}\exp\left[-\frac{(J+1/2)^2}{2\sigma^2(E_x)}\right].
\end{equation}
Here, $\sigma(E_x)$ is the excitation-energy dependent spin-cutoff parameter. 
Following the same line of arguments as in Ref.~\cite{Markova2022}, we choose the form of $\sigma(E_x)$ provided by Ref.~\cite{Gilbert65}:
\begin{equation}
\label{eq:6}
\sigma^2(S_n)=0.0888a\sqrt{\frac{S_n-E_1}{a}}A^{2/3},
\end{equation}
with the level density and back-shift parameters $a$ and $E_1$ obtained from the global parameterization of Ref.~\cite{Egidy05}. 
In line with the previously published results on $^{111-113,116,120,124}$Sn \cite{Markova2021,Markova2023}, the slopes of the NLDs and, therefore, the slopes of the GSFs obtained with this spin-cutoff parameter are in good agreement with the Coulomb excitation data for all isotopes \cite{Bassauer2020b}.
Moreover, recent calculations by Hilaire \textit{et al.}~\cite{Hilaire2023} within the quasiparticle random-phase approximation plus boson expansion reveal a smaller spin cutoff parameter than obtained, e.g., in the rigid-body moment of inertia approximation \cite{Egidy05}, which further supports the use of $\sigma(E_x)$ in Eq.~(\ref{eq:6}).

Combining Eq.~(\ref{eq:4}) and Eq.~(\ref{eq:5}), $\rho(S_n)$ takes the following form:
\begin{equation}
\label{eq:7}
\rho(S_n)=\frac{2\sigma^2}{D_0}\frac{1}{(J_t+1)\exp(-\frac{(J_t+1)^2}{2\sigma^2})+J_t\exp(-\frac{J_t^2}{2\sigma^2})}.
\end{equation}

For $^{111,112,122,124}$Sn no neutron resonance data are available, and the $\rho(S_n)$ values were estimated from  systematics 
in the same way as it was done in Ref.~\cite{Markova2023}. 
The slopes of the NLDs in $^{120,124}$Sn were additionally constrained with the shape method \cite{Wiedeking2020} as described in Refs.~\cite{Markova2021,Markova2022}.

\begin{table*}[t]
\caption{\label{tab:table_2}Parameters used for the normalization of the NLDs and GSFs for $^{111-113,116-122,124}$Sn.}
\begin{ruledtabular}
\begin{tabular}{lccccccccccccc}
Nucleus & $S_n$ & $D_0$ & $a$ & $E_1$ & $E_d$ & $\sigma_d$ & $\sigma(S_n)$ & $\rho(S_n)$ & $T$ & $E_0$ & $\langle\Gamma_{\gamma}\rangle$ \\ 
& (MeV) & (eV) & (MeV$^{-1}$) & (MeV) & (MeV) & & & ($ 10^5$ MeV $^{-1}$) &  (MeV) & (MeV) & (meV)\\
\noalign{\smallskip}\hline\noalign{\smallskip}
 $^{111}$Sn & 8.169 & 120(36)\footnotemark[1] & 12.05 & -0.29 & 1.08(7) & 2.7(4) & 4.6(5) & 3.54(127)\footnotemark[1] & 0.67$^{+0.03}_{-0.02}$ & -0.06$^{+0.04}_{-0.11}$  &  76(18)\footnotemark[1] \\ \noalign{\smallskip}
 
 $^{112}$Sn & 10.788  & 3(1)\footnotemark[1] & 12.53 & 1.12 & 2.83(4) & 2.8(4) & 4.8(5) & 24.61(80)\footnotemark[1] & 0.71$^{+0.02}_{-0.02}$ & 0.66$^{+0.09}_{-0.08}$ &  87(34)\footnotemark[2] \\ \noalign{\smallskip}
 
$^{113}$Sn & 7.744 & 172(10) & 12.77 & -0.27 & 1.88(2) & 3.5(7) & 4.6(5) & 2.50(51) & 0.63$^{+0.01}_{-0.01}$ & 0.20$^{+0.04}_{-0.04}$ &  73(8) \\ \noalign{\smallskip}

$^{116}$Sn & 9.563 & 55(5) & 13.66 & 1.13 & 2.27(6) & 2.7(5) & 4.8(5) & 4.28(91) & 0.79$^{+0.02}_{-0.02}$ & -0.50$^{+0.09}_{-0.04}$ &  118(10) \\ \noalign{\smallskip}

$^{117}$Sn & 6.943 & 507(60) & 13.62 & -0.21 & 1.11(11) & 2.5(2) & 4.6(5) & 0.85(19) & 0.69$^{+0.02}_{-0.02}$ & -0.57$^{+0.07}_{-0.05}$ &  53(3) \\ \noalign{\smallskip}

$^{118}$Sn & 9.326 & 61(7) & 13.94 & 1.14 & 2.48(4) & 2.7(5) & 4.8(5) & 3.89(87) & 0.76$^{+0.02}_{-0.02}$ & -0.18$^{+0.08}_{-0.13}$ &  117(20) \\ \noalign{\smallskip}

$^{119}$Sn & 6.483 & 700(100) & 13.80 & -0.30 & 1.32(2) & 3.7(10) & 4.6(5) & 0.61(15) & 0.69$^{+0.02}_{-0.02}$ & -0.80$^{+0.05}_{-0.12}$ &  45(7) \\ \noalign{\smallskip}

$^{120}$Sn & 9.105 & 95(14) & 13.92 & 1.12 & 2.53(4) & 3.7(5) & 4.8(5) & 2.49(60) & 0.75$^{+0.02}_{-0.02}$ & 0.07$^{+0.10}_{-0.05}$ &  121(25)\footnotemark[3] \\ \noalign{\smallskip}

$^{121}$Sn & 6.170 & 1485(130) & 13.63 & -0.39 & 1.26(5) & 4.0(8) & 4.5(5) & 0.28(6) & 0.70$^{+0.02}_{-0.02}$ & -0.70$^{+0.10}_{-0.09}$ &  36(3) \\ \noalign{\smallskip}

$^{122}$Sn & 8.815 & 95(28)\footnotemark[1] & 13.58 & 1.07 & 2.75(2) & 4.2(8) & 4.7(5) & 1.31(46)\footnotemark[1] & 0.76$^{+0.03}_{-0.04}$ & 0.07$^{+0.19}_{-0.06}$ &  87(19)\footnotemark[1] \\ \noalign{\smallskip}

$^{124}$Sn & 8.489 & 96(27)\footnotemark[1] & 12.92 & 1.03 & 2.77(3) & 3.3(5) & 4.7(5) & 0.87(26)\footnotemark[1] & 0.79$^{+0.02}_{-0.04}$ & -0.31$^{+0.16}_{-0.09}$ &  82(19)\footnotemark[1] \\ \noalign{\smallskip}

\end{tabular}
\end{ruledtabular}
\footnotetext[1]{From systematics.}
\footnotetext[2]{Modified (see Ref.~\cite{Markova2023}).}
\footnotetext[3]{Modified (see Ref.~\cite{Markova2022}).}
\end{table*}

The low-$\gamma$ energy boundary applied 
for the Oslo method analysis limits the experimental NLDs to energies $\approx 1-2$ MeV below the neutron threshold. 
To be able to connect the experimental NLD fixed to low-lying discrete states and the $\rho(S_n)$ data point, the experimental data were extrapolated with the constant-temperature model \cite{Ericson58, Ericson59, Gilbert65}:
\begin{equation}
\label{eq:8}
    \rho_{CT}(E_x)=\frac{1}{T_{{CT}}}\exp(\frac{E_x-E_0}{T_{{CT}}}),
\end{equation}
with the temperature $T$ and excitation-energy shift $E_0$ used as free fit parameters. 
As discussed earlier in Ref.~\cite{Guttormsen2015}, as well as in Refs.~\cite{Markova2022,Markova2023} specifically for the Sn isotopes, this model provides the best $\chi^2$ fit to the experimental data above $\approx 3$ MeV in the odd nuclei and $\approx 4$ MeV in the even-even nuclei. 

The uncertainty bands for the experimental NLDs comprise the statistical errors and systematic uncertainties due to the unfolding and the first-generation method, as outlined in Ref.~\cite{Schiller2000}. Analogous to the analysis of $^{111-113}$Sn \cite{Markova2023}, the experimental errors of $D_0$ were propagated together with the assumed additional 10\% errors for the $\sigma(S_n)$ parameter and added to the total uncertainty band in each case 
as described in Refs.~\cite{Kullmann2019,Larsen2023}. 
Following Ref.~\cite{Markova2023}, for the cases where the normalization input parameters are obtained from the systematics, a symmetric 30\% error for the extracted $D_0$ parameters was assumed and propagated in the total uncertainty bands. 
The choices of the errors for $\sigma(S_n)$ and $D_0$ extracted from the systematics are supported in all the studied cases by a good agreement of the slopes of the Oslo method GSFs with those extracted from the Coulomb excitation data \cite{Bassauer2020b}. 
All input values used for the normalization of the NLDs for $^{111-113,116-122,124}$Sn are provided in Table \ref{tab:table_2}.    
\begin{figure*}[t]
\includegraphics[width=1.0\linewidth]{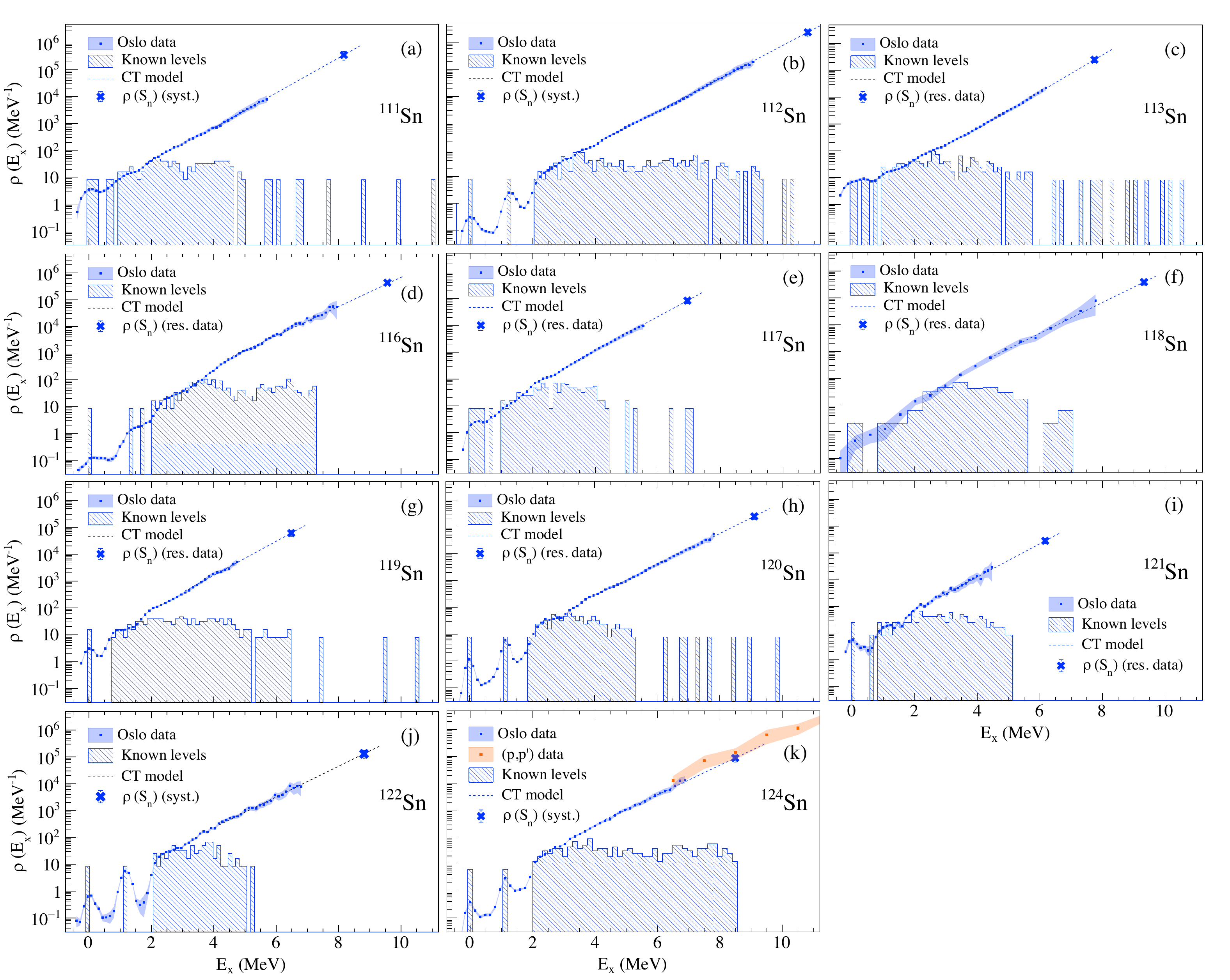}
\caption{\label{fig: LDs}
Experimental NLDs for $^{111-113,116-122,124}$Sn obtained with the Oslo method shown together with the $\rho(S_n)$ values and the constant-temperature fits. The orange band indicates the NLD of for $^{124}$Sn from the fluctuation analysis of the Coulomb excitation data \cite{Bassauer2020b}.
}
\end{figure*}

\subsubsection{\label{subsubsec 2.2.2: GSF normalization}Normalization of the GSFs}

The only remaining normalization parameter to be determined after the NLD normalization is the scaling $B$ of the experimental GSF. 
It can be obtained from the average total radiative width $\langle\Gamma_{\gamma}\rangle$ extracted in neutron-resonance studies \cite{Mughabghab18}. 
In general, the average total radiative width for excited states with spin $J$ and parity $\pi$ at excitation energy $E_x$ can be written as \cite{Kopecky1990}:
\begin{equation}
\label{eq:9}
\begin{split}
    \langle\Gamma(E_x, J, \pi)\rangle=&\frac{1}{2\pi\rho(E_x, J, \pi)}\sum_{XL}\sum_{J_f,\pi_f}\int_{E_{\gamma}=0}^{E_x}dE_{\gamma}\times\\&\times\mathcal{T}_{XL}(E_\gamma)\rho(E_x-E_{\gamma},J_f,\pi_f),
\end{split}
\end{equation}
where the $\gamma$-ray transmission coefficient ${T}_{XL}(E_\gamma)$ governs $\gamma$ decays of these states to final states with spins and parities $J_f^{\pi_f}$ with photons of type $X$ ($E$ and $M$ for the electric and magnetic character, respectively) and multipolarity $L$. Further, the transmission coefficient is linked to the GSF, $f_{XL}(E_\gamma)$, by the relation \cite{Kopecky1990}
\begin{equation}
\label{eq:10}
    {T}_{XL}(E_\gamma)=2\pi E^{2L+1}_\gamma f_{XL}(E_\gamma).
\end{equation}
Within the limits of excitation and $\gamma$-ray energies chosen for the Oslo method analysis, the $\gamma$ decay is expected to be dominated by dipole transitions of mixed $E$1+$M$1 nature (see e.g. \cite{Kopecky1990,Larsen2013}). 
Specifically, for the case of $s$-wave neutron capture on a target nucleus with ground-state spin-parity $J_t^{\pi_t}$ (where $J_t \neq 0)$, Eq.~(\ref{eq:9}) takes the following form:
\begin{align}
\label{eq:11}   
    \langle\Gamma_{\gamma}\rangle=&\langle\Gamma(S_n, J_t\pm1/2, \pi_t)\rangle=\frac{1}{2\rho(S_n, J_t\pm1/2, \pi_t)}\times \notag\\&\times\int_{E_{\gamma}=0}^{S_n}dE_{\gamma} E_{\gamma}^3f(E_\gamma)\rho(S_n-E_{\gamma})\times\notag\\&\times\sum_{J=-1}^{1}g(S_n-E_{\gamma},J_t\pm1/2+J)
\end{align}
with $1/\rho(S_n, J_t\pm1/2, \pi_t)=D_0$ and $g(E_x,J)$ is given by Eq.~(\ref{eq:5}). 
By analogy with Refs.~\cite{Markova2021,Markova2022,Markova2023}, the excitation-energy dependence of the spin-cutoff parameter is adopted from Ref.~\cite{Capote2009}:
\begin{equation}
\label{eq:12}  
    \sigma^2(E_x) = \sigma_d^2 + \frac{E_x-E_d}{S_n-E_d}[\sigma^2(S_n)-\sigma_d^2],
\end{equation}
which is additionally supported by microscopic calculations (see e.g. \cite{Uhrenholt13}). Here, $\sigma_d$ is the spin-cutoff determined at excitation energy $E_d$ from low-lying discrete states with definite spin and parity assignment, within the energy range where the level scheme can be considered complete. 

The average total radiative widths are available from neutron resonance experiments for most of the studied nuclei. 
For the lightest $^{111,112}$Sn isotopes, the values of $\langle\Gamma_{\gamma}\rangle$ and the corresponding uncertainties were extracted from the systematics in the same way as in Ref.~\cite{Markova2023}. 
For $^{124}$Sn and $^{122}$Sn, the $\langle\Gamma_{\gamma}\rangle$ values and their uncertainties were extracted from the same systematics according to the procedure described in Refs.~\cite{Markova2021,Markova2022}.

The total experimental uncertainty bands for the GSFs in all the studied Sn isotopes include the statistical errors, systematic uncertainties of the unfolding and the first-generation method combined with the propagated uncertainties due to the normalization input parameters $D_0$, $\sigma(S_n)$, $\sigma_d$, $E_d$, and $\langle\Gamma_{\gamma}\rangle$. 
All of the discussed parameters and uncertainties are presented in Table~\ref{tab:table_2}.

\section{\label{sec 3: Results}Experimental results}
\subsection{\label{subsec 3.1: NLDs}Nuclear level densities of S\MakeLowercase{n} isotopes}

The experimental 
NLDs of $^{111-113,116-122,124}$Sn extracted with the Oslo method 
are shown in Fig.~\ref{fig: LDs} together with the corresponding constant-temperature model fits and the $\rho(S_n)$ values. 
In all cases, the low-lying discrete 
states are well reproduced by the experimental results up to $\approx 2.5-3.5$ MeV in even-even and $\approx 1.5-2.5$ MeV in even-odd isotopes. 
Above these energies, the level schemes can no longer be considered complete, failing to follow the exponential increase observed in the Oslo data. 
For the experiments performed with the oldest configuration of the setup ($^{116,118}$Sn) the excitation-energy resolution is noticeably worse than in the most recent experiments with SiRi, and the ground and the first excited states are seen as broad bumps rather than well defined peaks as in, e.g., $^{120,124}$Sn. 
The ground state is somewhat underestimated in most of the cases as compared to the first excited state(s), which appears to be a commonly observed feature in OCL experiments (see e.g. the case of $^{46}$Ti \cite{Guttormsen2011}). 
This might be 
linked to the reaction mechanism favoring slightly higher spins of populated states \cite{Markova2023} or the structure of the states involved, hindering direct transitions from the quasi-continuum to the ground state.
This issue will be addressed in more detail in a forthcoming publication on $^{64}$Zn. 

At higher excitation energies, all NLDs are well described by the constant-temperature model. 
The new ($p,p^{\prime}\gamma)$ data on $^{117}$Sn perfectly reproduce the earlier published result using a $^3$He beam~\cite{Agvaanluvsan09-1}, while the new ($p,p^{\prime}\gamma)$ data on $^{119}$Sn do not seem to reveal the same step-like structures below $\approx 4$ MeV as seen in the previous experiment ~\cite{Toft2010}. 
The present result is more consistent with the NLDs of the neighboring $^{117,121}$Sn and reveals only one clear step-like structure in the vicinity of $\approx 2.6$ MeV. This feature is also seen in $^{117}$Sn (present work and Ref.~\cite{Agvaanluvsan09-1}) and $^{113}$Sn (Ref.~\cite{Markova2023}) and might be potentially linked to the first broken neutron pair. The features reported in the older experiment should be treated with care, considering the poor statistics of the $^{119}$Sn($^3$He,$\alpha\gamma$)$^{118}$Sn and $^{119}$Sn($^3$He,$^3$He$^{\prime}\gamma$)$^{119}$Sn experiments.

A comparison of the NLDs for $^{111,113}$Sn with the NLD for $^{115}$Sn from neutron-evaporation experiments \cite{Roy2021} has already been discussed in Ref.~\cite{Markova2023}. 
The density of 1$^{-}$ states for $^{124}$Sn from a fluctuation analysis of the Coulomb excitation data on $^{124}$Sn \cite{Bassauer2020b} was compared to the corresponding density from the Oslo data in Ref.~\cite{Markova2022}. 
In contrast to the present work, the latter publication presents the data normalized using a spin-cutoff parameter based on the rigid-body moment of inertia~\cite{Egidy05}, providing slightly steeper slopes of the NLDs (lower temperatures). 
When applied to all the studied isotopes, the slopes of the corresponding GSFs appear to be steeper than those from the Coulomb excitation data, unless corrected with the shape method \cite{Guttormsen2022,Markova2022}. 
Due to difficulties with the application of the shape method (such as insufficient statistics to provide reliable results at relatively high $\gamma$-ray energies), a consistent correction of the NLD slopes in the studied Sn nuclei is complicated. 
The spin-cutoff parameter from Eq.~(\ref{eq:6}) was found to provide the most consistent description of the NLDs and GSFs in all isotopes, supported by the good agreement with the Coulomb excitation data. 

It is important to check if the agreement of the fluctuation-analysis result and the Oslo data for $^{124}$Sn presented in Ref.~\cite{Markova2023} still holds for the new normalization approach. 
The spin distribution from Eq.~(\ref{eq:9}) was applied to the density of 1$^{-}$ states obtained with the fluctuation analysis, assuming an equal contribution of states with negative and positive parities above the lower limit of the analysis at $E_x\approx 6.5$ MeV. 
The resulting total NLD for the $^{124}$Sn isotope is shown together with the Oslo data in Fig.~\ref{fig: LDs}(k). This comparison of the total NLDs is almost identical to that for the $1^{-}$ states from Ref.~\cite{Markova2022}. Indeed, the rigid-body spin-cutoff parameter provides a broader spin distribution, predicting slightly steeper slopes of the total NLDs and, accordingly, smaller fractions of $1^{-}$ states. 
These two effects compensate each other as seen in the analogous figure of Ref.~\cite{Markova2022}. 
With the normalization approach of the present work, the total NLD obtained with the Oslo data lies closer to the bottom of the error band of the fluctuation-analysis data, staying within the band together with the $\rho(S_n)$ value. 
Hence, when applying the normalization procedure from Sec.~\ref{subsubsec 2.2.1: NLD normalization} for a consistent description of all isotopes in the present work, the main conclusions of Ref.~\cite{Markova2022} still hold. 

\begin{figure}[t]
\includegraphics[width=1.0\linewidth]{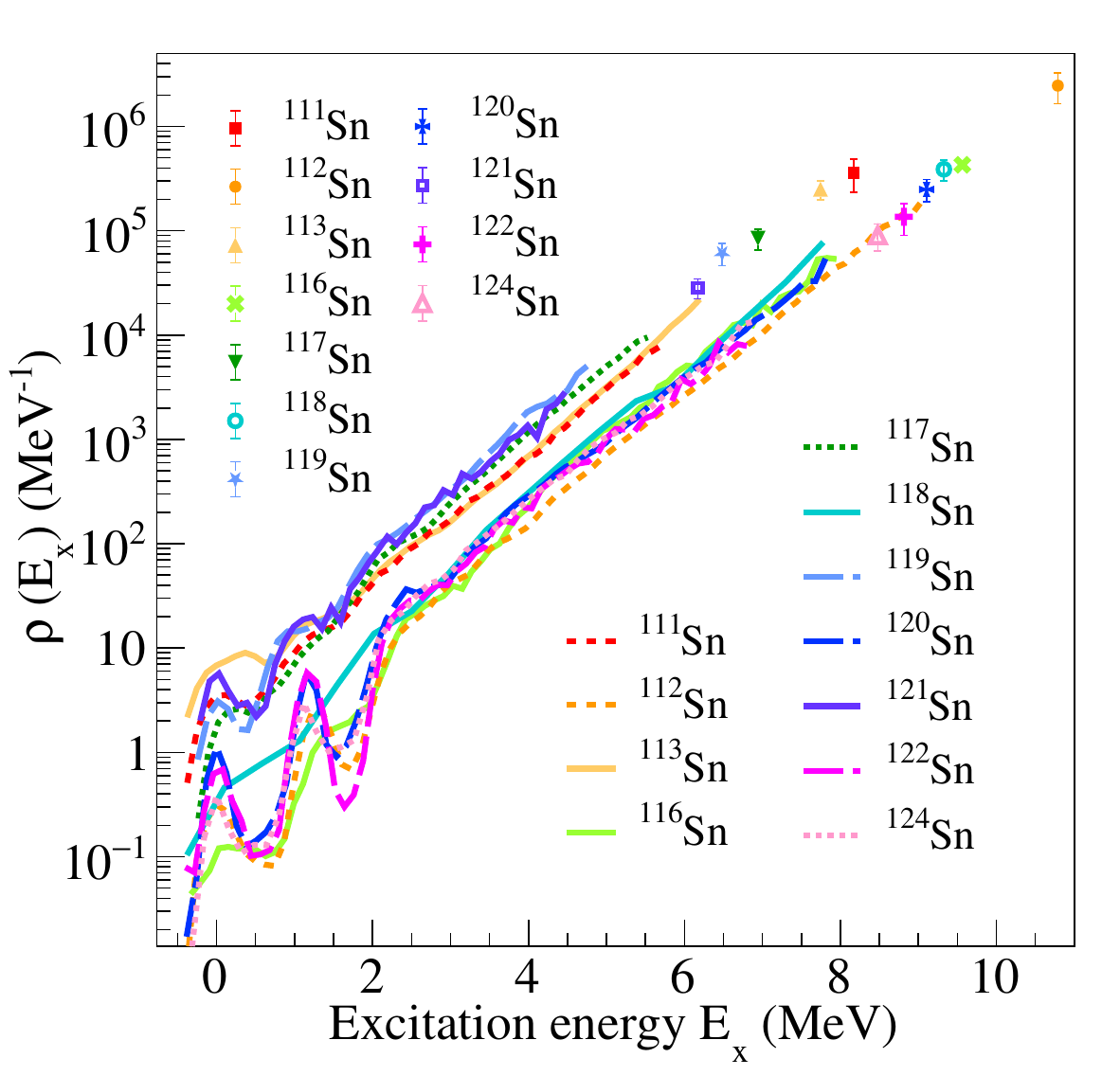}
\caption{\label{fig: all LDs}
Experimental NLDs for $^{111-113,116-122,124}$Sn obtained with the Oslo method. The uncertainty bands are omitted for enhancing the clarity of the figure.
}
\end{figure}

The NLDs of all the studied isotopes are shown together in Fig.~\ref{fig: all LDs}. 
With the same approach to the normalization, the slopes of all NLDs appear to be very similar, corresponding to  temperatures of $T\approx 0.6-0.8$ MeV. 
For the even-even nuclei, the NLDs show a good agreement in absolute values, well within the uncertainty bands 
from $\approx 2$ MeV up to the neutron separation energies. 
This is expected for the even-even Sn isotopes considering their similar structural properties. 
A much better agreement between the NLDs of the even-odd nuclei from $\approx 1.5$ MeV to $\approx 4$ MeV is achieved with the present consistent normalization approach, in contrast to the comparison with the earlier published data presented in Ref.~\cite{Markova2022}. 
As expected, the level densities of the odd nuclei are systematically higher than those of the even-even ones due to the unpaired neutron. 
The NLDs of the lightest $^{112}$Sn and $^{111,113}$Sn are slightly lower when compared to the heavier even-even and even-odd isotopes, respectively. 
This is similar to a pattern observed for the lightest Ni isotopes in Ref.~\cite{Ingeberg2022}.
However, the NLDs of the heavier Sn isotopes display a smaller spread than the heaviest studied Ni isotopes in \cite{Ingeberg2022}, and the trend of the NLD increasing with neutron number as discussed in \cite{Ingeberg2022} is not apparent in the Sn data. 

\subsection{\label{subsec 3.2: GSFs and PDRs} Experimental low-lying electric dipole strength in Sn isotopes and its evolution}

\begin{figure*}[h]
\includegraphics[width=1.0\linewidth]{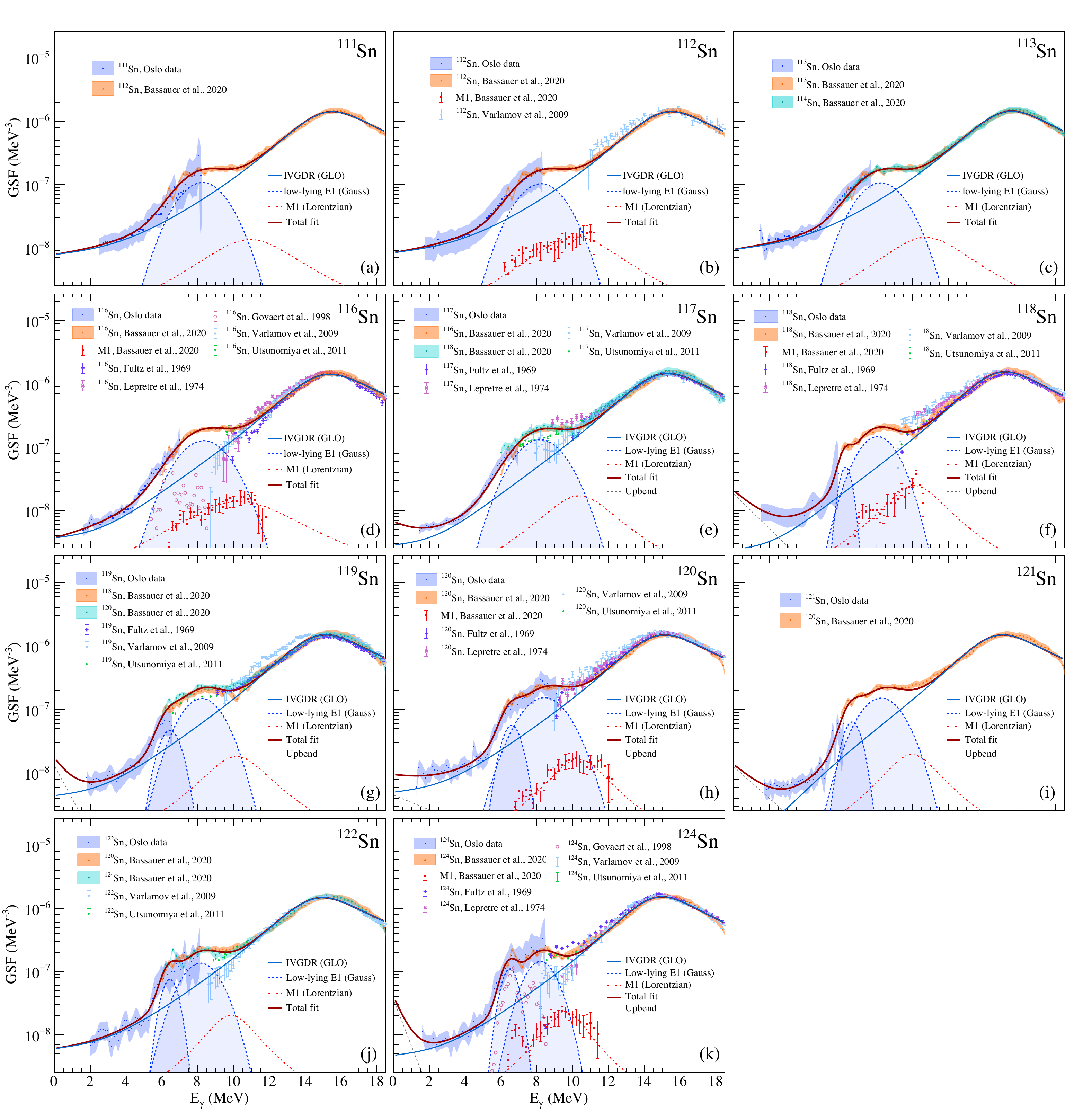}
\caption{\label{fig: Sn GSFs}
Experimental GSFs of $^{111}$Sn (a), $^{112}$Sn (b), $^{113}$Sn (c), $^{116}$Sn (d), $^{117}$Sn (e), $^{118}$Sn (f), $^{119}$Sn (g), $^{120}$Sn (h), $^{121}$Sn (i), $^{122}$Sn (j), and $^{124}$Sn (k) shown together with the $(p,p^{\prime})$ data from Ref.~\cite{Bassauer2020b} and the ($\gamma, n$) experimental data by Varlamov \textit{et al.} \cite{Varlamov2009}, Fultz \textit{et al.} \cite{Fultz69}, Lepr\^{e}tre \textit{et al.} \cite{Lepretre74}, Utsunomiya \textit{et al.} \cite{Utsunomiya2009,Utsunomiya2011}, and Govaert \textit{et al.} \cite{Govaert98}. 
The total fits of the experimental data are shown as solid magenta lines and the fits of the IVGDR as solid blue lines. The low-lying $E1$ components, obtained from fits with Eq.~(\ref{eq:16}), are shown as shaded light-blue areas. The $M1$ data from the Coulomb excitation experiment \cite{Bassauer2020b} are shown for $^{112,116,118,120,124}$Sn with corresponding Lorentzian fits (dashed red lines).}
\end{figure*}

The  GSFs 
normalized as described in section \ref{subsubsec 2.2.2: GSF normalization} are shown together with the ($p,p^{\prime}$) and $(\gamma,n$) data in Fig.~\ref{fig: Sn GSFs}. 
The Coulomb excitation strengths are available only for the even-even $^{112,114,116,118,120,124}$Sn isotopes, while photoabsorption data from experiments performed in Saclay \cite{Fultz69}, Livermore \cite{Lepretre74}, and Moscow \cite{Varlamov2009} are also available for the even-odd stable Sn targets. 
The photoabsorption cross sections provided by these experiments cover a wide range above the neutron threshold. 
Overall, these data agree quite well with the ($p,p^{\prime}$) data in the vicinity of the IVGDR peak at $\approx 15$ MeV (for a more detailed discussion see Ref.~\cite{Bassauer2020b}). 
The largest deviations of the $(\gamma,n$) data from each other and the ($p,p^{\prime}$) experiments occur in the vicinity of the neutron threshold, which makes a consistent comparison with the Oslo data difficult. 
Because the Coulomb excitation strengths are available for lower energies (down to $\approx 6$ MeV), there is a sufficient overlap with the Oslo method GSFs in most cases. 
Therefore, we put greater emphasis on the ($p,p^{\prime}$) data when comparing with the Oslo method results than the $(\gamma,n$) data. 
In contrast to the above-mentioned $(\gamma,n$) experiments with quite large uncertainties close to the $S_n$ energy, the most recent $(\gamma,n)$ experiments on $^{116-120,122,124}$Sn using quasi-monoenergetic photon beams from laser Compton backscattering demonstrate a very good agreement with the ($p,p^{\prime}$) strengths 
where the data overlap ($\approx 6-12$ MeV).

In Fig.~\ref{fig: Sn GSFs}(a,b,c), the extracted GSFs for $^{112-113}$Sn are shown together with the ($p,p^{\prime}$) strength of $^{112}$Sn. 
In all three cases, the strengths agree well within the uncertainty bands as previously discussed in Ref.~\cite{Markova2023}. 
The ($p,p^{\prime}$) strength of $^{114}$Sn is in good agreement in slope and absolute value with both the Oslo method GSF of $^{113}$Sn and the ($p,p^{\prime}$) strength of $^{112}$Sn. 
This is also the case for $^{117}$Sn and $^{119}$Sn, shown together with the ($p,p^{\prime}$) strengths of the closest neighboring even-even $^{116,118}$Sn and $^{118,120}$Sn, respectively. 
The Oslo strengths of $^{116,120,124}$Sn agree exceptionally well within the uncertainty bands with the corresponding ($p,p^{\prime}$) strengths, as reported in Ref.~\cite{Markova2021}.  
Due to some minor updates in the response functions used for the unfolding since the time of the latter publication, the slope of the $^{124}$Sn is slightly steeper than the one reported in \cite{Markova2021}, while still being well within both systematic and statistical uncertainty bands of the earlier published strength. 
Thus, the conclusions of \cite{Markova2021} remain unchanged with the updated version of the GSF for $^{124}$Sn. 

Above $\approx 6$ MeV, the statistics of the older experiment on $^{118}$Sn are insufficient to draw any reliable conclusions on the low-lying dipole strength at these energies based on the Oslo data alone. For this reason, the Oslo method GSF of $^{118}$Sn is shown only up to $\approx 6$ MeV with the ($p,p^{\prime}$) strength being complementary at higher energies. 
The Oslo strength in this case agrees fairly well with the strengths of the neighboring even-even isotopes. 
Similarly, the $^{122}$Sn strength is shown up to $\approx 8$ MeV, being in good agreement with the ($p,p^{\prime}$) GSFs of $^{120}$Sn and $^{124}$Sn. 
No experimental data on $^{121}$Sn above the neutron threshold are available. 
Assuming a smooth evolution of the IVGDR strength with increasing neutron number as demonstrated by the Coulomb excitation results and the ($\gamma,n$) data, the Oslo strength was compared with the ($p,p^{\prime}$) GSF of the closest even-even $^{120}$Sn in Fig.~\ref{fig: Sn GSFs}(i). 
Similarly to $^{118}$Sn, both GSFs are in fair agreement in absolute values. 
Overall, the Oslo data reveal a smooth evolution of the low-lying dipole strength below the neutron threshold, with neighboring isotopes having similar shapes of the GSF, consistent with the observed strength in the IVGDR region.

\subsection{Empirical model fits to the data}
To address the evolution of the low-lying $E1$ strength with neutron number, it should be consistently extracted from the total dipole response in the studied nuclei. 
As mentioned earlier, the Oslo method does not distinguish between $E1$ and $M1$ components. 
Therefore, other experimental constraints on the $M1$ spin-flip resonance are highly desired. 
In the case of the Sn isotopes, the experimental magnetic dipole strengths are available from the multipole decomposition analysis of the Coulomb excitation data for even-even Sn isotopes \cite{Bassauer2020b}.
These data provide sufficient information to extract systematics, which allow us to estimate the $M1$ component in the neighboring nuclei. 
To determine the contribution of the LEDR (or PDR in other works, e.g. \cite{Toft2011}), the low-energy tail of the IVGDR has to be subtracted from the remaining total $E1$ response. 
Unfortunately, there is no common approach to predict the PDR strength distribution in nuclei, nor any consensus on how to separate it from the IVGDR strength. 
The experimental strength distributions obtained with complementary probes for the same nucleus, albeit being insightful from a nuclear-structure perspective, do not suggest any consistent, quantitative answer to this problem. 
One of the frequently adopted approaches to extract the LEDR is to assume a model for the IVGDR and estimate the remaining LEDR by subtracting the tail of the modeled IVGDR from the total $E1$ response. 
Alternatively, a model is assumed to reproduce the general shape of the LEDR yielding the best fit to the experimental data. 
This technique is often used for the interpretation of experimental strength distributions in neutron-rich nuclei \cite{Adrich2005,Rossi2013} or analyses featuring Oslo-type experiments (see e.g. \cite{Guttormsen2022,Larsen2023}). 
Furthermore, the total theoretical or experimental $E1$ strength can be summed up to a chosen threshold, or within a certain energy range with no assumptions made regarding the tail of the IVGDR and its contribution (see, e.g., \cite{Paar2007,Govaert98,Krumbholz2015, Muscher2020}). 

In this section, we exploit the first of the two above-mentioned methods, with as few assumptions as possible, to quantify the evolution of the LEDR in the Sn isotopes from the Oslo results and Coulomb excitation data within the $\simeq 2-18$ MeV $\gamma$ energy range. 
In accordance with Ref.~\cite{Markova2023}, we choose the enhanced Generalized Lorentzian model (GLO) to describe the IVGDR data \cite{Capote2009}:
\begin{equation}
\label{eq:13}  
    \begin{split}
        f_{E1}(E_{\gamma})=&\frac{1}{3\pi^2\hbar^2 c^2}\sigma_{E1}\Gamma_{E1}\times\\
        &\times \biggl [ E_{\gamma} \frac{\Gamma_{\rm KMF}(E_{\gamma},T_f)}{(E_{\gamma}^2-E_{E1}^2)^2+E_{\gamma}^2\Gamma_{\rm KMF}^2(E_{\gamma}, T_f)}+\\
        &+0.7\frac{\Gamma_{\rm KMF}(E_{\gamma}=0,T_f)}{E_{E1}^3}\biggr ],
    \end{split}
\end{equation}
with $E_{E1}$, $\Gamma_{E1}$, $\sigma_{E1}$ being the IVGDR centroid energy, width, and cross section, respectively. 
The $\Gamma_{KMF}$ width corresponds to a temperature-dependent width in the Kadmenskii-Markushev-Furman model \cite{Kadmenskii1983}:
\begin{equation}
\label{eq:14}  
    \Gamma_{\rm KMF}(E_{\gamma},T_f)=\frac{\Gamma_{E1}}{E_{E1}^2}(E_{\gamma}^2+4\pi^2T_f^2),
\end{equation}
where $T_f$ is the temperature of the final states.
\begin{table*}[t]
\caption{\label{tab:table_3}Parameters used for the description of the IVGDR and the $M1$ strength in the studied Sn isotopes.}
\begin{ruledtabular}
\begin{tabular}{lccccccccc}
\noalign{\smallskip}
 Nucl. & $E_{E1}$ & $\Gamma_{E1}$ & $\sigma_{E1}$ & $T_{f}$ & $E_{M1}$ & $\Gamma_{M1}$ & $\sigma_{M1}$ & $C_{\rm up}$ & $\eta_{\rm up}$\\
 & (MeV) & (MeV) & (mb) & (MeV) & (MeV) & (MeV) & (mb) & (10$^{-8}$ MeV$^{-3}$) & (MeV$^{-1}$)\\
\noalign{\smallskip}\hline\noalign{\smallskip}\noalign{\smallskip}
 $^{111}$Sn & 16.15(9)  & 5.49(31) & 264.5(93) & 0.67(4) & 11.22(24)\footnotemark[1] & 5.15(39)\footnotemark[1] & 1.73(15)\footnotemark[1] & -- & --\\ \noalign{\smallskip}
 $^{112}$Sn & 16.14(9)  & 5.46(31) & 265.9(95) & 0.70(5) & 10.45(43) & 4.77(53) & 1.77(21) &  -- &  --\\ \noalign{\smallskip}
 $^{113}$Sn & 16.14(6)  & 5.25(23) & 274.4(74) & 0.75(3) & 10.99(20)\footnotemark[1] & 4.72(32)\footnotemark[1] & 1.84(11)\footnotemark[1] & -- & --\\ \noalign{\smallskip}
 $^{116}$Sn & 16.09(10)  & 6.03(35) & 251.3(91) & 0.43(2) & 10.79(41) & 6.28(96) & 1.70(13) & -- & -- \\ \noalign{\smallskip}
 $^{117}$Sn & 15.98(7)  & 5.84(26) & 257.1(74) & 0.38(6) & 10.54(12)\footnotemark[1] & 3.86(20)\footnotemark[1] & 2.05(7)\footnotemark[1] & 0.38(10) & 0.59(9)\\ \noalign{\smallskip}
 $^{118}$Sn & 15.78(10)  & 5.50(47) & 270.0(153) & 0.35(7) & 10.26(19) & 3.21(34) & 2.90(24) & 1.84(56) & 0.63(8)\\ \noalign{\smallskip}
 $^{119}$Sn & 15.82(6)  & 5.77(22) & 264.0(69) & 0.45(9) & 10.31(9)\footnotemark[1] & 3.44(16)\footnotemark[1] & 2.16(8)\footnotemark[1] & 1.29(15) & 1.20(31)\\ \noalign{\smallskip}
 $^{120}$Sn & 15.82(9)  & 5.79(39) & 262.8(111) & 0.48(14) & 10.45(18) & 3.13(33) & 1.97(16) & 0.45(7) & 0.27(8)\\ \noalign{\smallskip}
 $^{121}$Sn & 15.72(6)  & 5.86(24) & 255.3(66) & 0.22(12) & 10.08(9)\footnotemark[1] & 3.01(14)\footnotemark[1] & 2.27(11)\footnotemark[1] & 1.27(63) & 0.58(22)\\ \noalign{\smallskip}
 $^{122}$Sn & 15.67(3)  & 5.85(11) & 258.7(24) & 0.52(3) & 9.97(10)\footnotemark[1] & 2.79(15)\footnotemark[1] & 2.32(13)\footnotemark[1] & -- & -- \\ \noalign{\smallskip}
 $^{124}$Sn & 15.59(7)  & 5.37(28) & 266.8(90) & 0.49(4) & 9.66(14) & 2.42(20) & 2.61(16) & 3.52(115) & 1.67(40)\\ \noalign{\smallskip}
\end{tabular}
\footnotetext[1]{From systematics.}
\end{ruledtabular}
\end{table*}

The standard Lorentzian function, commonly used to fit the photo-neutron cross section above the neutron threshold (see e.g. \cite{Krumbholz2015}), is known to overshoot the low-energy flank of the strength and is, therefore, excluded from  consideration here. 
Among other phenomenological models, the Generalized Fermi Liquid model by Mughabghab~\cite{Mughabghab2000} and the Hybrid model by Goriely~\cite{Goriely98} are either able to capture the strength distribution at low energies ($\approx 2-4$ MeV) and fail to follow the left flank of the IVGDR or vice versa, being more appropriate in cases with a less steep GSF below $S_n$ (e.g. \cite{Larsen2023}). 
The Simplified Modified Lorentzian function (SMLO) \cite{Plujko2002} results in a milder overshoot below the threshold energy as compared to the standard Lorentzian, while still failing to reproduce the low-energy tail of the Oslo data. 
Microscopic strength distributions provided by calculations within Skyrme-Hartree-Fock with Bardeen-Cooper-Schrieffer pairing~\cite{Goriely2002}, Skyrme-Hartree-Fock-Bogoliubov \cite{Goriely2004}, Gogny-Hartree-Fock-Bogoliubov \cite{Goriely2018} with the quasiparticle random-phase approximation (QRPA) and its temperature-dependent extension \cite{Daoutidis2012} require certain modifications (scaling of the absolute value or width, and often an energy shift) to be able to reproduce the IVGDR part. 
Even with these modifications, the microscopic calculations still can not be used to extract the LEDR consistently in all the studied Sn isotopes due to an overshoot at low $\gamma$ energies in some cases. 
However, the GLO model is sufficiently flexible within a relatively wide energy range to obtain simultaneously a satisfactory fit of the IVGDR peak and the tail of the strength at $\approx 2-4$ MeV.

For a consistent modeling of the $M1$ part in all Sn isotopes, we assume a simple Lorentzian shape of the spin-flip resonance:
\begin{equation}
\label{eq:15}  
    f_{M1}(E_{\gamma})= \frac{1}{3\pi^2\hbar^2c^2}\frac{\sigma_{M1}\Gamma_{M1}^2E_{\gamma}}{(E_{\gamma}^2-E_{M1}^2)^2+E_{\gamma}^2\Gamma_{M1}^2}
\end{equation}
with centroid energy $E_{M1}$, maximum cross section $\sigma_{M1}$, and width $\Gamma_{M1}$. 
The experimental $M1$ strengths from the $(p,p^{\prime})$ experiments on even-even Sn isotopes are quite fragmented, and the Lorentzian function merely reproduces the overall shapes and total integrated $M1$ strengths. 
As the contribution of this component to the total GSF is less than 10\% at the maximum, the details of this fit are of little influence on the final results.

The LEDR component was parameterized with Gaussian peaks:
\begin{equation}
\label{eq:16}  
    f_{E1}^{\rm low}(E_{\gamma})= C_{E1}^{\rm low}\frac{1}{\sqrt{2\pi}\sigma_{E1}^{\rm low}}\exp[-\frac{(E_{\gamma}-E_{E1}^{\rm low})^2}{2\sigma_{E1}^{\rm low}}],
\end{equation}
with centroid $E_{E1}^{\rm low}$, width $\sigma_{E1}^{\rm low}$, and absolute value $C_{E1}^{\rm low}$. 
The choice of this fit function is not immediately obvious, and for more moderate slopes of the GSFs, the LEDR is also well reproduced by one or a combination of several Lorentzian peaks \cite{Guttormsen2022}. 
For the Sn isotopes having steep slopes at $E_\gamma \approx 4-6$ MeV and relatively flat strength distributions at lower energies, the best fits to the experimental data in the energy range up to $S_n$ are achieved using Gaussian peaks. 
The Gaussian model was also applied to reproduce the LEDR in very neutron-rich nuclei \cite{Adrich2005, Rossi2013}.

To account for the flat low-energy tails of the GSF, we follow the prescription of Ref.~\cite{Schwengner2013} suggesting an exponential form of the upbend feature based on the comparison of shell-model calculations and experimental data on Zr and Mo isotopes:
\begin{equation}
\label{eq:16}  
    f_{\rm up}(E_{\gamma})= C_{\rm up}\exp(-\eta_{\rm up}E_{\gamma}),
\end{equation}
with scaling and slope parameters $C_{\rm up}$ and $\eta_{\rm up}$. 
The Oslo data on the Sn isotopes reveal no clear sign of a strong upbend at low $E_{\gamma}$, but show rather flat strength distributions. 
Since the data are restricted to $E_{\gamma}\gtrsim 2$ MeV, we do not have sufficient information to reveal any reliable systematics on the upbend as was done recently for the Nd isotopes \cite{Guttormsen2022}. 
In the present work, the upbend is treated solely as a fit component at low $\gamma$-ray energies, having negligible impact on the extracted LEDR.

\begin{table*}[]
\caption{\label{tab:table_4}Parameters used for the description of the low-lying $E1$ strengths, integrated low-lying $E1$ strengths, and the corresponding exhausted fractions of the TRK sum rule in the studied Sn isotopes.}
\begin{ruledtabular}
\begin{tabular}{lccccccccc}
\noalign{\smallskip}
Nucl. & \multicolumn{3}{c}{Peak 1} & & \multicolumn{3}{c}{Peak 2} & &\\ \noalign{\smallskip}
\cline{2-4} \cline{6-8} \noalign{\smallskip}
& $E_{E1}^{\rm low}$ & $\sigma_{E1}^{\rm low}$ & $C_{E1}^{\rm low}$ & & $E_{E1}^{\rm low}$ & $\sigma_{\rm E1}^{low}$ & $C_{E1}^{\rm low}$ & Integrated strength & TRK \\ 
& (MeV) & (MeV) & ($10^{-7}$ MeV$^{-2}$) & & (MeV) & (MeV) & ($10^{-7}$ MeV$^{-2}$) & (MeV mb) & (\%) \\
\noalign{\smallskip}\hline\noalign{\smallskip} \noalign{\smallskip}
$^{111}$Sn & -- & -- & -- & & 8.26(9)  & 1.23(7)  & 3.32(23) & 31.6(25) & 1.92(15) \\ \noalign{\smallskip}
$^{112}$Sn & --  & -- & -- & & 8.24(9)  & 1.22(8) & 3.17(24) & 30.1(25) & 1.81(15) \\ \noalign{\smallskip}
$^{113}$Sn & --  & -- & -- & & 8.23(6)  & 1.23(6) & 3.27(17) & 31.1(18) & 1.86(11) \\ \noalign{\smallskip}
$^{116}$Sn & --  & -- & -- & & 8.33(8)  & 1.29(6) & 4.08(25) & 39.2(28) & 2.29(16) \\ \noalign{\smallskip}
$^{117}$Sn & --  & -- & -- & & 8.18(6)  & 1.26(5) & 4.15(19) & 39.2(20) & 2.28(12) \\ \noalign{\smallskip}
$^{118}$Sn & 6.27(18)  & 0.33(10) & 0.40(15) & & 8.04(21)  & 1.00(20) & 3.71(65) & 37.3(55) & 2.16(32) \\ \noalign{\smallskip}
$^{119}$Sn & 6.44(13)  & 0.56(11) & 0.67(35) & & 8.23(11)  & 1.06(11) & 3.97(41) & 42.7(40) & 2.45(23) \\ \noalign{\smallskip}
$^{120}$Sn & 6.59(11)  & 0.50(9) & 0.71(22) & & 8.42(12)  & 1.19(10) & 4.60(43) & 50.1(47) & 2.86(27) \\ \noalign{\smallskip}
$^{121}$Sn & 6.62(9)  & 0.48(7) & 0.77(20) & & 8.25(9)  & 1.11(7) & 4.20(31) & 45.9(33) & 2.61(19) \\ \noalign{\smallskip}
$^{122}$Sn & 6.45(5)  & 0.43(5) & 0.82(16) & & 8.17(7)  & 1.00(7) & 3.40(20) & 38.1(21) & 2.15(12) \\ \noalign{\smallskip}
$^{124}$Sn & 6.49(5)  & 0.43(5) & 1.20(22) & & 8.20(7)  & 0.83(12) & 2.99(34) & 37.3(36) & 2.08(20) \\ \noalign{\smallskip}

\end{tabular}
\end{ruledtabular}
\end{table*}

To disentangle the $E1$ and $M1$ components of the total GSF, we first fit the $M1$ strength distributions for $^{112,114,116,118,120,124}$Sn and build the systematics for the parameters of the Lorentzian functions to reconstruct the $M1$ part in the even-odd isotopes and $^{122}$Sn. 
The strength distribution in the neighboring even-odd nuclei can be expected to be even more fragmented, but the total amount of the $M1$ strength should still be close to that in the even-even neighbors. 
Further, the total $E1+M1$ strength is fitted with the combined function $f_{tot}=f_{E1}+f_{E1}^{\rm low}+f_{M1}+f_{\rm up}$, where the parameters of $f_{M1}$ are kept constant. 
This corresponds to a simultaneous fit of all $E1$ features of the total strength. 
Alternatively, the IVGDR region can be fitted first with the $f_{E1}$ function (e.g. see Refs.~\cite{Pogliano2023_1, Markova2023}), then keeping its parameters constant while constraining the remaining LEDR component. 
The latter method yields slightly larger $\chi^2$ values than the simultaneous fit. 
Since both methods provide integrated strengths well in agreement within the error bars, we are limiting the analysis to the simultaneous fit of the total $E1$ strength with $f_{tot}$. 
The data to be fitted are the Oslo method GSFs and the corresponding $(p,p^{\prime})$ strengths for even-even $^{112,116,118,120,124}$Sn. 
For $^{122}$Sn and the odd isotopes $^{111,113,117,119,121}$Sn, the Coulomb excitation data for the closest even-even isotopes were used ($^{120,124}$Sn, $^{112}$Sn, $^{112,114}$Sn, $^{116,118}$Sn, $^{118,120}$Sn, and $^{120}$Sn, respectively). 
As the ($p,p^{\prime}$) data demonstrate the same smooth evolution of the IVGDR with increasing neutron number as the $(\gamma,n)$ data, while also being more consistent in the vicinity of $S_n$, they were preferred over the $(\gamma,n)$ strengths for all the considered odd isotopes.

The ($p,p^{\prime}$) data have been reported to reveal a peak-like structure at $\approx 6.4-6.5$ MeV in all studied even-even isotopes \cite{Bassauer2020b}.
This feature becomes especially prominent in $^{124}$Sn. 
The lack of data points at energies below $\approx 6$ MeV did not allow to perform a fit of this feature by using the Coulomb excitation data alone. 
In general, the energy resolution in Oslo-type of experiments and relatively large systematic uncertainties make it difficult to observe such features in the Oslo data. 
However, some hints of a peak-like feature at $\approx 6.4-6.5$ MeV in the $^{124}$Sn Oslo strength.
Also, the Oslo method GSFs become gradually steeper for heavier Sn isotopes, allowing for introducing an additional Gaussian peak to the fit of the LEDR of the heavier isotopes starting from $^{118}$Sn. 
Indeed, a double-peaked LEDR yields an improved fit of the experimental data between $\approx 5$ and 11 MeV for these isotopes as compared to a single Gaussian peak. 
For the lighter Sn isotopes, the second peak is not well defined and, therefore, was not included in the total fit. 
All the above-mentioned fit parameters of the IVGDR, the $M1$, and the upbend functions are presented in Table~\ref{tab:table_3}. 
The characteristics of the extracted LEDR components are shown in Table~\ref{tab:table_4}.

\begin{figure}[t]
\includegraphics[width=1.0\linewidth]{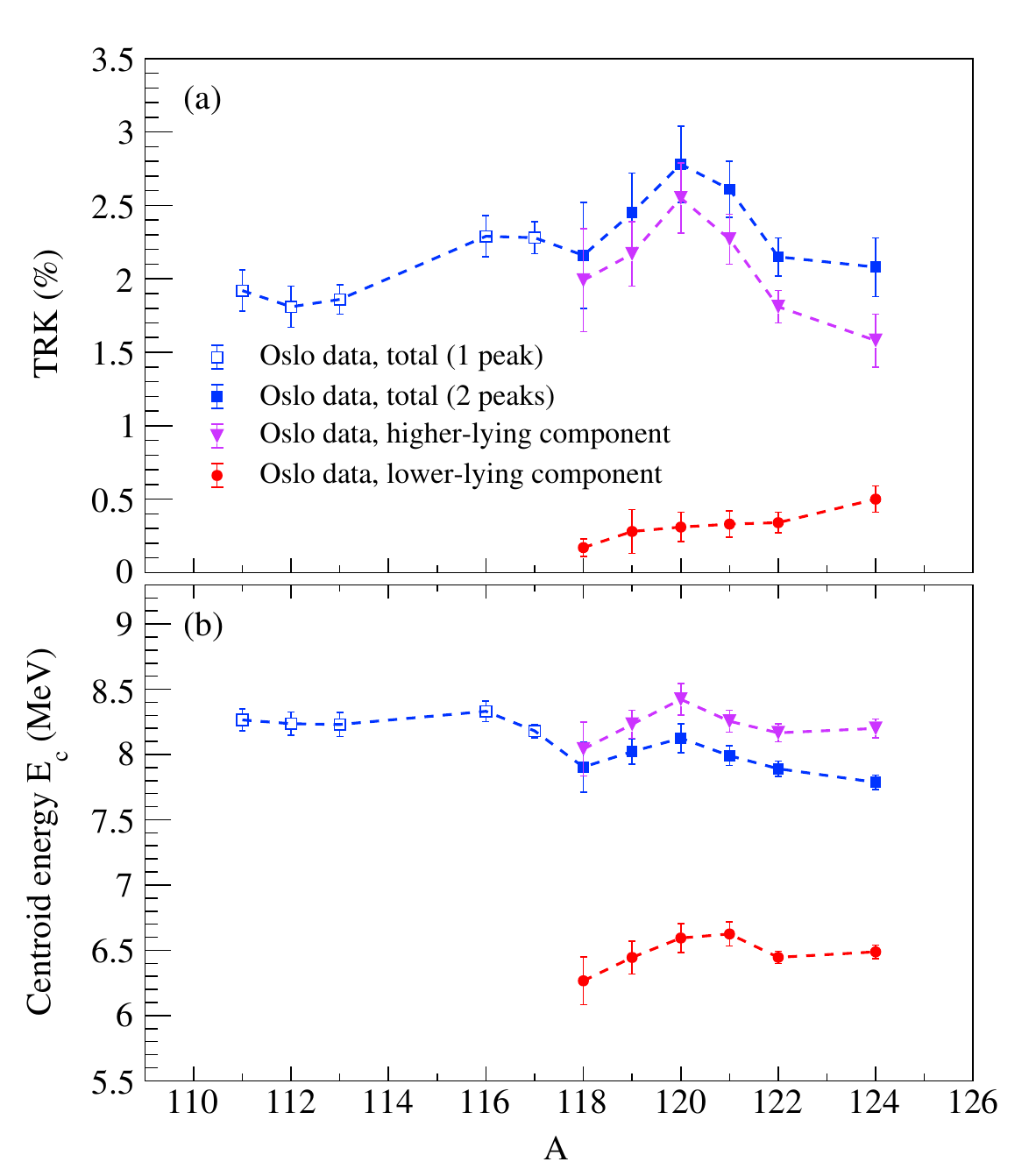}
\caption{\label{fig: RDR exp syst}
TRK values (a) and energy centroids (b) for the total extracted LEDR in Sn isotopes, its lower-lying, and higher-lying components. Hollow squares correspond to a single Gaussian peak fit, filled squares correspond to the sum (a) and strength-averaged centroids (b) of two Gaussian peaks.
}
\end{figure}

The systematics of the total integrated strength 
of the LEDR in the studied Sn isotopes in terms of the exhausted fraction of the
TRK sum rule \cite{Thomas1925,Reiche1925,Kuhn1925} are shown in Fig.~\ref{fig: RDR exp syst}(a). 
For the isotopes fitted with double peaks, this fraction is also shown for the smaller, low-lying and the larger, higher-lying components separately. 
The LEDR extracted according to the above-mentioned procedure appears to correspond to $\approx 2-3$\% of the TRK sum rule 
for all the Sn nuclei considered here. 
No clear systematic increase of the total strength with increasing neutron number is observed. 
On the contrary, the energy-weighted integrated strengths are quite similar for all the studied nuclei within the uncertainties, peaking around $^{120}$Sn ($\approx 3$\% of the TRK sum rule). 
The Oslo GSF for $^{120}$Sn (normalized independently of the Coulomb excitation data) is quite close within the uncertainty bands to the GSFs of the neighboring isotopes. 
The appearance of the local maximum is mainly driven by the $(p,p^{\prime})$ data which show slightly larger absolute values in the energy range from 8 to 10 MeV than in the other isotopes.

Considering that the GSFs of the even-even isotopes were used for constraining the LEDR in the even-odd isotopes, all of them, as expected, reveal a somewhat averaged behavior with respect to the even-even neighbors. 
This is additionally supported by the Oslo GSFs, demonstrating no clear odd-even effects and showing a smooth trend from the lightest to the heaviest nuclei. 

\subsection{Discussion and comparison with theoretical predictions in the literature}

Most of the theoretical approaches 
predict that the low-lying $E1$ strength should increase with the proton-neutron asymmetry parameter, while also not being a function of the neutron excess alone. 
How steep and monotonous this trend is strongly depends on the theoretical framework and the criteria applied to identify the potential PDR (or LEDR in general) strength.
Specifically for the Sn isotopes, the microscopic relativistic quasiparticle random-phase approximation (RQRPA) and relativistic quasiparticle time-blocking approximation (RQTBA) calculations of Ref.~\cite{Litvinova2009}, Hartree–Fock–Bogoliubov plus quasiparticle phonon model (QPM) calculations of Ref.~\cite{Tsoneva2008}, and the study based on the Vlasov equation approach of Ref.~\cite{Baran2014} demonstrate a smooth general increase of the LEDR strength with increasing neutron excess. 
Provided the experimental constraints shown in Fig.~\ref{fig: RDR exp syst}(a), no claims on any strong dependence of the integrated strength on increasing neutron number between $^{111}$Sn and $^{124}$Sn can be made. A weak dependence, if present, is obscured by a local peak in strength around $^{120}$Sn. 
Exploiting a single-peak fit for $^{118-122,124}$Sn affects neither the general trend nor the absolute values of the integrated strength within the limits of the estimated uncertainties. 
The choice of the fit for the $M1$ component was also found to have negligible impact on the obtained values. 
For example, a more detailed fit of the $M1$ strength with three and two Lorentzian functions for $^{120}$Sn and $^{124}$Sn, respectively, results only in $\approx 1\%$ reduction of the values shown in Fig.~\ref{fig: RDR exp syst}(a).

A similar local maximum of the integrated strength in the vicinity of $^{120}$Sn has previously been observed within the random-phase approximation (RPA) approach \cite{Piekarewicz2006}, interrupting an almost linear correlation of the integrated PDR strength and the neutron skin thickness. 
This effect was related to a gradual filling of the $1h_{11/2}$ neutron orbital in the heavier isotopes, suppressing transitions of low multipolarity within the PDR region. 
Furthermore, pairing correlation effects were included in relativistic Hartree–Bogoliubov (RHB) plus RQRPA calculations~\cite{Paar2007}, which revealed a somewhat similar local peak at $^{120-124}$Sn in the strength evolution.
This was attributed to an interplay between reduced pairing correlations and shell effects when approaching the $N=82$ shell closure. 
A similar pattern emerges in calculations from a relatively recent study on the isovector and isoscalar response in Sn nuclei within the time-dependent Hartree-Fock (TDHF) approach \cite{Burrello2019}. 
The open-shell nucleus $^{120}$Sn was shown to have a larger fraction of the energy-weighted sum rule exhausted within the PDR region in both the isoscalar and isovector channels, as compared to the doubly magic $^{100}$Sn and $^{132}$Sn. 
Among the studied TDHF density profiles, $^{120}$Sn appears to exhibit a slightly more diffuse surface, potentially correlated with the enhancement of the strength in this nucleus. 
All of these studies employ an upper limit for the extraction of the total integrated strength, which complicates a direct quantitative comparison with the present experimental results. 
We note that the 
local maximum of the strength at $^{120}$Sn is a subtle feature considering the uncertainties in the data. 
However, a theoretical interpretation would still be important, in particular whether it presents a local feature based on shell structure or a general phenomenon in nuclei with sufficient neutron excess. 
A possible link to the explanations offered in Refs.~\cite{Paar2007,Burrello2019} requires further investigations.

It is interesting to note that the energy-weighted integrated strength of the smaller, low-lying component in $^{118-122,124}$Sn increases approximately linear with neutron number [see Fig.~\ref{fig: RDR exp syst}(a)]. 
As mentioned earlier, the feature at $\approx6.5$ MeV appearing in the ($p,p^{\prime}$) strength in all the studied Sn isotopes has, indeed, been noted to become more prominent toward $^{124}$Sn \cite{BassauerThesis}. 
Nevertheless, this trend is quite subtle, and the total exhausted strength of this peak-like structure is limited to only 0.1-0.5\% of the TRK sum rule.  
A similar concentration of the isoscalar strength between 5.5 and 7 MeV has been observed earlier in the studies of $^{124}$Sn with the ($\alpha,\alpha^{\prime}\gamma)$ \cite{Endres2012} and ($^{17}$O,$^{17}$O$^{\prime}\gamma$) reactions \cite{Pellegri2014}. 
Combined with the $(p,p^{\prime})$ and $(\gamma,\gamma^{\prime})$ data, they provide experimental evidence of a structural splitting of the LEDR in this nucleus into a group of lower-lying states of mixed isovector-isoscalar nature, observed in all the mentioned probes, and higher-lying states of isovector nature, seen only in the $(p,p^{\prime})$ and $(\gamma,\gamma^{\prime})$ experiments. 
The correspondence with the isoscalar probes and large ground state branching ratios observed in ($\gamma,\gamma^\prime)$ experiments suggest that the lower-energy peak represents the isovector response of the PDR. The implications of this result will be further discussed in Ref.~\cite{PLB}.

The employed Gaussian fit allows to easily access the evolution of the centroid of the LEDR with increasing neutron number, presented in Fig.~\ref{fig: RDR exp syst}(b).
For $^{118-122,124}$Sn, the centroids of both components are shown together with the strength-weighted average centroid for the total LEDR.
The strength in all the studied isotopes is concentrated at $\approx 7.8-8.3$ MeV, while the lower and the higher peaks in $^{118-122,124}$Sn show almost constant centroid energies of $\approx 6.4-6.5$ and $\approx 8.2$ MeV, respectively.
Provided that the centroids of the lower and higher components are almost unchanged, a mild decrease of the total LEDR centroid reflects the same strength redistribution as in Fig.~\ref{fig: RDR exp syst}(a), with gradually more strength grouped at $\approx 6.4-6.5$ MeV toward $^{124}$Sn. 
The decrease of the LEDR/PDR centroid is reproduced in RQRPA \cite{Paar2007, Litvinova2009} and QPM \cite{Tsoneva2008} calculations for Sn isotopes, appearing also in isotopic chains of other elements~\cite{Tertychny2007,Paar2007}. 
The observed experimental trend of Fig.~\ref{fig: RDR exp syst}(b) contradicts the previously extracted Oslo systematics of the LEDR centroids in Ref.~\cite{Toft2011}.
This is mainly due to the great inconsistency of the photoabsorption data close to the neutron threshold, which the previous fits in Ref.~\cite{Toft2011} heavily relied on.

\section{\label{sec 4: Exp plus theory} Comparison with ab initio-based model calculations}

Nuclear response theory is the most practical tool to quantify the nuclear strength functions for a wide energy range. 
At the simplest level, the response theory is confined by the 
RPA or its superfluid extension, QRPA. 
Using Feynman diagrams, (Q)RPA is represented by a one-loop diagram of the two-fermion in-medium propagator, while in the most fundamental \textit{ab initio} equation-of-motion (EOM) framework \cite{AdachiSchuck1989, DukelskyRoepkeSchuck1998}, QRPA is obtained by neglecting two-particle-two-hole ($2p2h$) and higher-rank correlations in the interaction kernel.  
In the EOM of Rowe \cite{Rowe1968}, (Q)RPA is associated with the simplest one-particle-one-hole  (two-quasiparticle) $1p1h$ ($2q$) excitation operator, which generates the excited states by its action on a Hartree-Fock (Hartree-Fock-Bogoliubov) ground state.
(Q)RPA is known to reproduce the basic properties of giant resonances and soft modes; however, it fails at describing fine spectral details. 
More accurate solutions involve higher complexity ($npnh$) correlations in both the excited states and the ground state of the nucleus.
 
All approximations beyond (Q)RPA are derivable from the \textit{ab initio} EOM for the two-fermion response function \cite{LitvinovaSchuck2019,Litvinova2022} by retaining more complex correlations, in particular, in the dynamical kernel.
The leading approximation beyond (Q)RPA contains the quasiparticle-vibration coupling (qPVC) in the minimal coupling scheme, which includes $2q\otimes \rm \rm phonon$ configurations in the intermediate two-fermion propagator. 
The vibrations (phonons) emerge naturally as correlated $2q$ pairs, with the qPVC vertices as the new order parameters. 
This approach admits realistic implementations that employ effective interactions adjusted in the framework of density functional theory. 
With such interactions, reasonable phonons can be obtained already within (Q)RPA, and the qPVC can then be combined with subtraction restoring the self-consistency of the \textit{ab initio} framework \cite{Tselyaev2013}. 

The first self-consistent microscopic approach, which includes qPVC in terms of $2q\otimes \rm phonon$ configurations, was presented in Ref.~\cite{LitvinovaRingTselyaev2008} and applied to the dipole response of medium-heavy nuclei. 
This implementation was a major step toward as a universal theory of nuclear structure rooted in particle physics, named relativistic nuclear field theory, and used the effective meson-exchange interaction \cite{Lalazissis1997,NL3star}. 
The approach \cite{LitvinovaRingTselyaev2008} to the response function was based on a phenomenological assumption about the leading role of $2q\otimes \rm phonon$ configurations and the time-blocking technique \cite{Tselyaev1989}; thus identified as the relativistic quasiparticle time-blocking approximation (RQTBA). 
In Refs.~\cite{LitvinovaSchuck2019,Litvinova2022} the complete response theory was obtained via \textit{ab initio} EOMs, where both the phenomenological qPVC and time blocking are ruled out as unnecessary ingredients.  
The relativistic EOM confined by the $2q\otimes \rm phonon$ (REOM$^2$) configurations with the QRPA phonons was shown to be essentially equivalent to RQTBA. 
However, in contrast to the phenomenological approach, REOM is an \textit{ab initio} theory extendable to configurations of arbitrary complexity. 
An example of such an extension was presented as REOM$^3$ accommodating $2q\otimes 2\rm phonon$ configurations in Refs.~\cite{LitvinovaSchuck2019,Litvinova2023a}.

\begin{figure}[t]
\includegraphics[width=1.0\columnwidth]{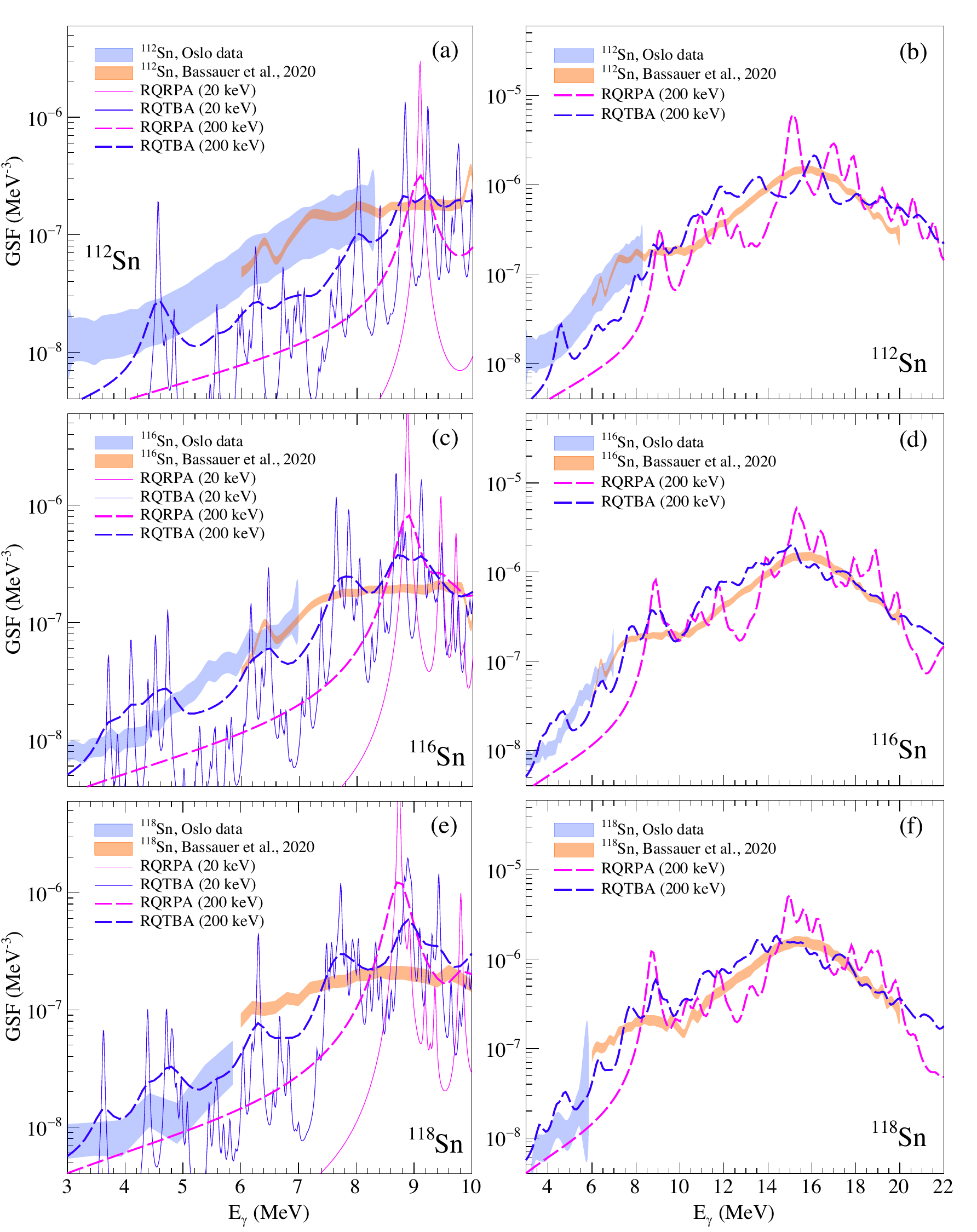}
\caption{\label{fig: RQRPA + RQTBA + Exp 1}
Calculated dipole strengths for $^{112,116,118}$Sn. The low-lying $E1$ transitions computed with the 20-keV (thin solid line) and 200-keV (thick dashed line) artificial widths are shown up to 10 MeV (a, c, e). The strengths computed with the 200-keV artificial width are also shown up to 22 MeV in (b, d, f). The blue and orange bands indicate the corresponding Oslo and $(p,p^{\prime})$ data. Calculations within the RQRPA and the RQTBA are shown with magenta and violet lines, respectively.
}
\end{figure}

\begin{figure}[t]
\includegraphics[width=1.0\columnwidth]{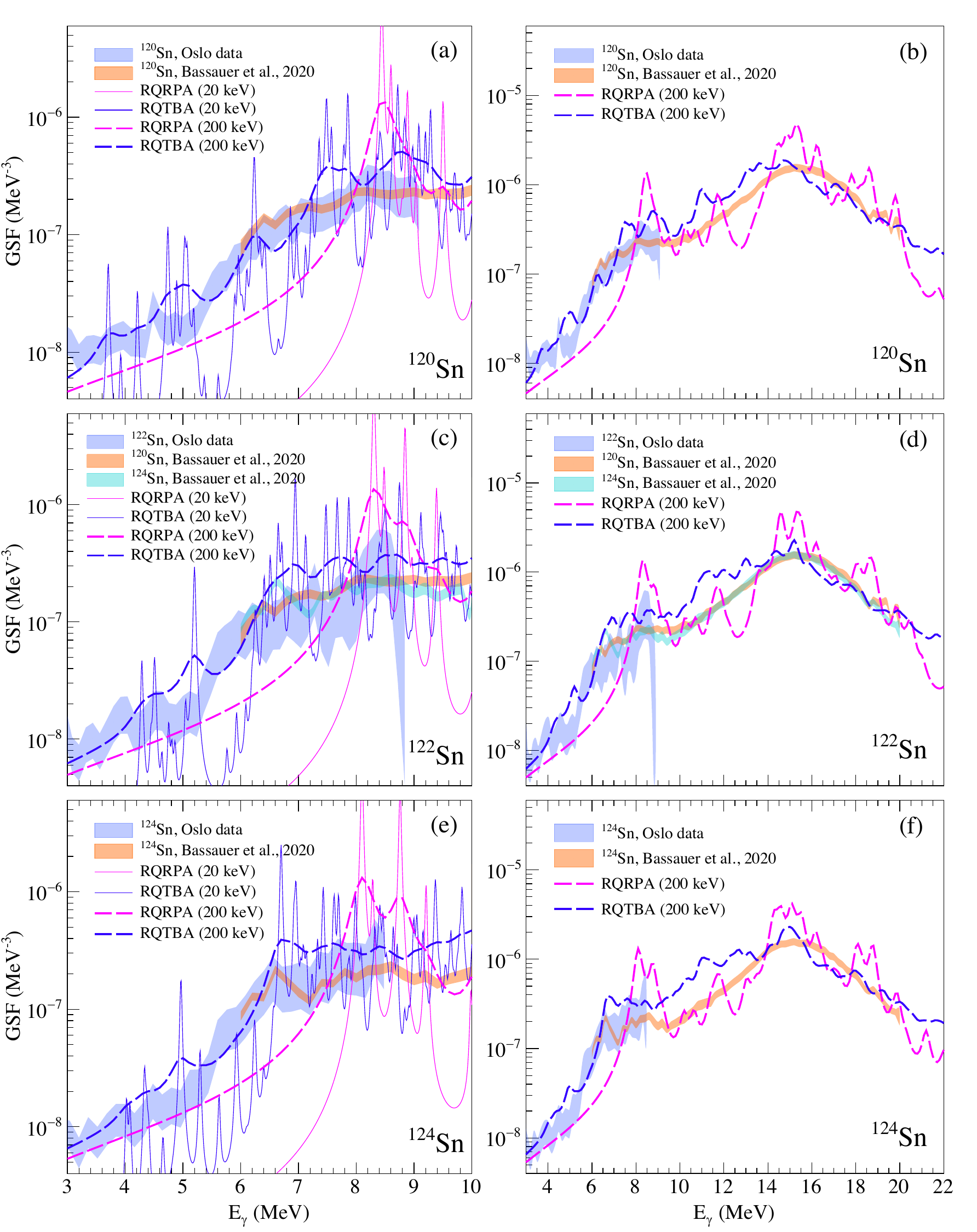}
\caption{\label{fig: RQRPA + RQTBA + Exp 2}
Same as in Fig.~\ref{fig: RQRPA + RQTBA + Exp 1}, but for $^{120,122,124}$Sn. For $^{122}$Sn, both $^{120}$Sn and $^{124}$Sn $(p,p^{\prime})$ data are shown.
}
\end{figure}

In this work, REOM$^2$-RQTBA was applied to calculations of the dipole response of the Sn isotopes under study in a broad energy range up to 25 MeV. 
The obtained strength distributions are compared to those of relativistic QRPA (RQRPA), which is used as a reference case, and to experimental data, as displayed in Figs.~\ref{fig: RQRPA + RQTBA + Exp 1} and \ref{fig: RQRPA + RQTBA + Exp 2}.  
The NL3* meson-exchange interaction \cite{NL3star} was employed in both approaches and the subtraction \cite{Tselyaev2013} is implemented in REOM$^2$.
In the latter, natural-parity phonons up to 15 MeV with $J = [1,6]$ and reduced transition probabilities above 5\% of the maximal one for each multipolarity were comprised in the intermediate $2q\otimes \rm phonon$ propagators.
The $2q$ configurations were included up to 100 MeV, while the $2q\otimes \rm phonon$ ones were accommodated up to 25 MeV.  
Calculations with two values of the smearing parameter $\Delta$, which is defined below, $\Delta$ = 20 keV and $\Delta$ = 200 keV are presented.


It is clearly seen from Figs. \ref{fig: RQRPA + RQTBA + Exp 1} and \ref{fig: RQRPA + RQTBA + Exp 2} that adding $2q\otimes \rm phonon$ configurations in RQTBA significantly changes the strength distribution, as compared to RQRPA.
Overall, the gross structures of the strength become fragmented and a significant portion moves toward lower transition energies. 
In particular, the PDR region below 10 MeV manifests considerable structural differences between the RQTBA and RQRPA approaches. 
Thus, the spreading of the IVGDR and the PDR structure occur mainly due to these configurations. 
In the paradigm of a self-consistent covariant many-body theory, its only input is the local meson-exchange interaction between two nucleons, while all the in-medium many-body correlations are included without changing the parameters of this interaction, or introducing new ones. 
Within this paradigm, the RQTBA strength distribution is a result of the fragmentation of the RQRPA modes. 

This can be understood from the general model-independent relationships, where the strength function $S(\omega)$ for a given energy (or frequency) $\omega$ is defined by Fermi's golden rule:
\bea
S(\omega) = \sum\limits_{\nu>0} \Bigl[ |\langle \nu|F^{\dagger}|0\rangle |^2\delta(\omega-\omega_{\nu}) - |\langle \nu|F|0\rangle |^2\delta(\omega+\omega_{\nu})
\Bigr], \nonumber\\
\label{SF}
\eea
where the summation is performed over all excited states $|\nu\rangle$ with transition energy $\omega_\nu = E_\nu - E_0$ with $E_0$ being the ground-state energy. 
The transition matrix element $\langle \nu|F^{\dagger}|0\rangle$ for the typical one-body external field operator
\be
\langle \nu|F^{\dagger}|0\rangle = \sum\limits_{12}\langle \nu|F_{12}^{\ast}\psi^{\dagger}_2\psi_1|0\rangle = \sum\limits_{12}F_{12}^{\ast}\rho_{21}^{\nu\ast},
\label{Frho}
\ee
is expressed via the transition densities
\be
\rho^{\nu}_{12} = \langle 0|\psi^{\dagger}_2\psi_1|\nu \rangle , 
\label{trden}
\ee
which are the weights of the pure particle-hole configurations $\psi^{\dagger}_2\psi_1$ in the single-particle basis $\{1\}$, on top of the ground state $|0\rangle$, in the excited states $|\nu\rangle$, and $\psi_1$ and $\psi_1^{\dagger}$ are the nucleonic field operators. 
Conventionally, the $\delta$ functions in Eq. (\ref{SF}) are represented by the Lorentz distribution
\be
\delta(\omega-\omega_{\nu}) = \frac{1}{\pi}\lim\limits_{\Delta \to 0}\frac{\Delta}{(\omega - \omega_{\nu})^2 + \Delta^2},
\ee
so that
\be
S(\omega) 
= -\frac{1}{\pi}\lim\limits_{\Delta \to 0} {\Im} \Pi(\omega+ i\Delta),
\label{SFDelta} 
\ee
where $\Pi(\omega)$ is the polarizability of the nucleus:
\be
\Pi(\omega) 
=  \sum\limits_{\nu} \Bigl[ \frac{B_{\nu}}{\omega - \omega_{\nu}}
- \frac{{\bar B}_{\nu}}{\omega + \omega_{\nu} }
\Bigr]
\label{Polar}
\ee
related to the transition probabilities $B_{\nu}$ and ${\bar B}_{\nu}$ of absorption and emission, respectively:
\be
B_{\nu} = |\langle \nu|F^{\dagger}|0\rangle |^2\ \ \ \ \ \ \ \ 
{\bar B}_{\nu} = |\langle \nu|F|0\rangle |^2
\label{Prob}.
\ee
Therefore, the strength function that quantifies the nuclear response to the given external field operator $F$ reads:
\be
S_F(\omega) = -\frac{1}{\pi}\lim_{\Delta\to 0}\Im\sum\limits_{121'2'}F_{12}R_{12,1'2'}(\omega+i\Delta)F^{\ast}_{1'2'},
\label{SFF}
\ee
where the central role in characterizing the nuclear structure is played by the response function $R_{12,1'2'}(\omega)$, whose spectral representation is:
\be
R_{12,1'2'}(\omega) = \sum\limits_{\nu>0}\Bigl[ \frac{\rho^{\nu}_{21}\rho^{\nu\ast}_{2'1'}}{\omega - \omega_{\nu} + i\delta} -  \frac{\rho^{\nu\ast}_{12}\rho^{\nu}_{1'2'}}{\omega + \omega_{\nu} - i\delta}\Bigr].
\label{respspec}
\ee
The poles of $R_{12,1'2'}(\omega)$ are at the energies $\omega_{\nu} = E_{\nu} - E_0$ of the excited states with respect to the ground state energy and $\delta \to +0$. 

Eq.~(\ref{respspec}) is the Fourier transform of the particle-hole propagator in a correlated medium:
\be
R_{12,1'2'}(t-t') =  -i\langle T\psi^{\dagger}(1)\psi(2)\psi^{\dagger}(2')\psi(1')\rangle,
\label{phresp}
\ee
where $\langle...\rangle$ is a shorthand notation for the expectation value in the ground state and $\psi(1), \psi^{\dagger}(1)$ are the fermionic field operators in the Heisenberg picture:
\bea
\psi(1) \equiv \psi_1(t_1) \equiv {e}^{iHt_1}\psi_1 {e}^{-iHt_1} \nonumber \\ 
\psi^{\dagger}(1) \equiv \psi^{\dagger}_1(t_1) \equiv {e}^{iHt_1}\psi^{\dagger}_1 {e}^{-iHt_1}.
\label{t-fields}
\eea 
where $t_1 = t_2 = t$, $t_{1'} = t_{2'} = t'$, and the fermionic Hamiltonian
\be
H =  \sum_{12}h_{12}\psi^{\dag}_1\psi_2 + \frac{1}{4}\sum\limits_{1234}{\bar v}_{1234}{\psi^{\dagger}}_1{\psi^{\dagger}}_2\psi_4\psi_3
\label{Hamiltonian1}
\ee
is specified by its one-body $h_{12}$ and two-body ${\bar v}_{1234}$ matrix elements.

The strength distribution for the given external field operator, which, in this work, is the electric dipole operator ($\mu$ denotes the magnetic substate)
\bea
F^{(E1)}_{1\mu} &=& \frac{eN}{A}\sum\limits_{i=1}^Z r_iY_{1\mu}({\hat{\bf r}}_i) - \frac{eZ}{A}\sum\limits_{i=1}^N r_iY_{1\mu}({\hat{\bf r}}_i),
\nonumber\\
\eea
is completely determined by the response function given by Eq.~(\ref{respspec}). 
The number of peaks in the resulting spectrum is equal to the number of terms in Eq.~(\ref{respspec}), and this number as well as the locations of the poles and transition densities are determined by the correlation content of the theory. 
The response function can be found from the Bethe-Salpeter-Dyson equation (BSDE), that is, in the operator form,
\be
R(\omega) = R^0(\omega) + R^0(\omega)\Bigl( K^0 + K^r(\omega)\Bigr)R(\omega),
\label{BSDE}
\ee
where $R^0(\omega)$ is the non-interacting particle-hole propagator in the mean field and the specific forms of the interaction kernels are given, for instance, in Refs. \cite{LitvinovaRingTselyaev2008,Litvinova2022}. For the application discussed in this work, it is essential that the RQRPA strength is obtained by neglecting completely the $K^r(\omega)$ term, which contains the qPVC correlations and is retained in RQTBA.

In the implementations using physical effective interactions, these interactions play the role of the static kernel $K^0$, which is the only interaction term in the (R)QRPA. 
In this approach, the dipole spectrum is characterized by two pronounced peaks, one at higher energy and the other at low energy, and a few less prominent structures. 
The high-energy peak is associated with the IVGDR formed by the out-of-phase oscillations of the proton and neutron Fermi liquids against each other, which follows from the radial behavior of the transition densities for this excitation. 
The two-quasiparticle content of the transition densities shows a rather high degree of collectivity when many $2q$ configurations contribute coherently to the probabilities (\ref{Prob}). 
The low-energy peak, often assigned as the PDR,  can be identified by similar means. 
Within this framework, it shows up as a neutron excess oscillation against the isospin-saturated core, also with some sign of collectivity \cite{Paar2007,Vretenar2012}. 

In comparison to data, the position of the main IVGDR peak is typically described reasonably well in (R)QRPA, however, this is often not the case for the PDR. 
The reason becomes evident only when going beyond the simplistic QRPA and including the frequency-dependent kernel $K^r(\omega)$ in the BSDE (Eq.~(\ref{BSDE})). 
The leading  $2q\otimes \rm phonon$ configurations included in REOM$^2$-RQTBA induce a similar fragmentation of both major peaks, and the resulting fragments overlap in the energy region between the IVGDR and PDR modes. This further leads to the same problem of separating the PDR contribution from the low-energy tail of the IVGDR as discussed earlier in Sec.~\ref{subsec 3.2: GSFs and PDRs}.

According to Eq.~(\ref{respspec}), adding $2q\otimes \rm phonon$ configurations produces additional terms in the sum on the right-hand side, i.e., more states in the resulting spectrum than with $2q$ configurations alone.
This is reflected in Figs.~\ref{fig: RQRPA + RQTBA + Exp 1} and \ref{fig: RQRPA + RQTBA + Exp 2} for all the isotopes under study.
The number of additional states is equal to the number of possible $2q\otimes \rm phonon$ combinations compatible with angular-momentum conservation. 
Overall, adding complex configurations leads to a better description of the data, in particular, the IVGDR width and the PDR fine structure. 
The importance of these configurations is especially evident when comparing the strength distributions at low energies shown in the left panels of Figs.~\ref{fig: RQRPA + RQTBA + Exp 1} and \ref{fig: RQRPA + RQTBA + Exp 2} obtained with the different values of the smearing parameter $\Delta$. 
It can be seen, for instance, that the low-energy portion of the RQRPA strength function is the purely artificial tail of the states located at 8-9 MeV. 
The finite strength below that energy originates solely from the smearing. 
Accordingly, the strength varies considerably when varying $\Delta$. 
In contrast, the RQTBA strength below the neutron threshold is essentially physical and the choice of the smearing parameter plays a minor role. 
This choice in the low level-density regime is stipulated by the experimental energy resolution, the finite level lifetime and missing higher-complexity configurations. 
In the context of this work, $\Delta = 200$ keV is an appropriate value, and calculations with $\Delta = 20$ keV are given to illustrate the fine structure of the strength. 
Nevertheless, the choice of this parameter plays a minor role as long as complex configurations are taken into account, which emphasizes the importance of these configurations for an adequate description of the low-energy nuclear strength functions.

\begin{figure}[t]
  \subfigure{\label{subfig 1: Theor + Exp syst}}
  \subfigure{\label{subfig 2: Theor + Exp syst}}
  \subfigure{\label{subfig 3: Theor + Exp syst}}
\includegraphics[width=1.0\columnwidth]{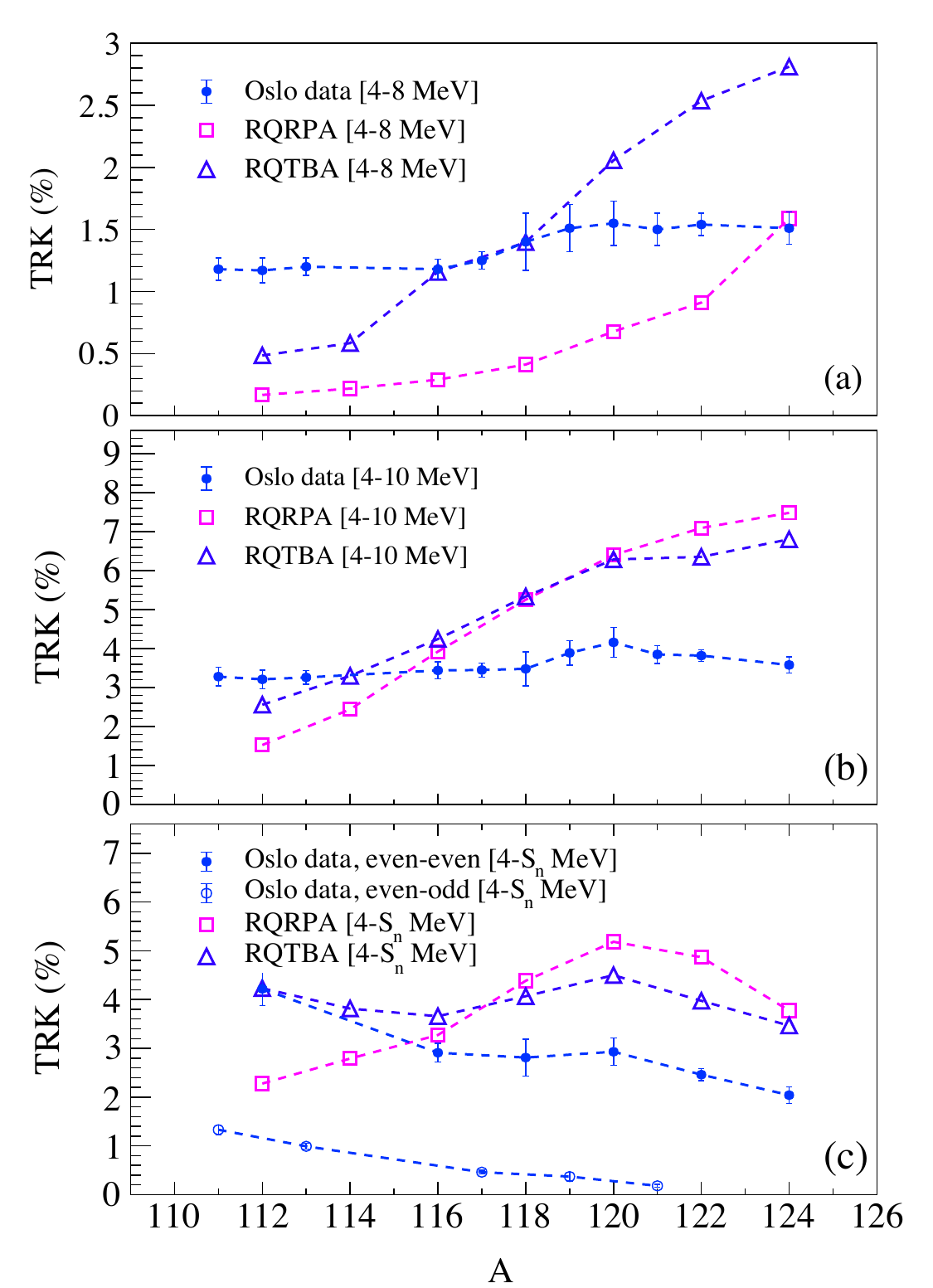}
\caption{\label{fig: Theor + Exp syst}
The evolution of the energy-weighted electric dipole strength extracted from the RQRPA and RQTBA calculations and the combined experimental Oslo and $(p,p^{\prime})$, integrated from 4 MeV up to 8 MeV (a), 10 MeV (b), and $S_n$ (c).
}
\end{figure}

To compare the overall behavior of the RQRPA, RQTBA, and experimental strength distributions on an equal footing, we extract the energy-weighted sums of the total electric dipole strength (the $M1$ component was subtracted from the experimental strength functions) within three energy ranges, namely $4-8$ MeV, $4-10$ MeV, and $4-S_n$, similarly to how it was done in Ref.~\cite{Litvinova2009}. 
This procedure corresponds to the second method of quantifying the LEDR contribution mentioned in the previous section. 
The extracted fractions of the TRK sum rule exhausted in each case are shown in Fig.~\ref{fig: Theor + Exp syst}. As discussed earlier, the experimental LEDR appears to be concentrated in the vicinity of 8 MeV and, naturally, a large fraction of the strength is located above this threshold. 
The experimental strength between 4 and 8 MeV corresponds to $\approx 1.2-1.5\%$ of the TRK sum rule. The monotonous increase of the RQRPA strength in this energy range is solely due to the artificial tails of the states at 8-9 MeV with the applied 200-keV smearing parameter, while the RQTBA strength increases gradually from $\approx 0.5-2.7\%$ of the TRK sum rule. 
Within the energy range up to 10 MeV, including most of the LEDR in the studied nuclei, the experimental strength exhausts $\approx 3-4\%$ of the TRK sum rule (here the IVGDR tail is included in the sum), in contrast to both RQRPA and RQTBA predicting a steady, monotonous increase of strength up to $\approx 7\%$ in $^{124}$Sn. 
Indeed, both approaches result in larger concentrations of strength in the immediate vicinity of 8-10 MeV as compared to the experimental strength distribution, gradually increasing with neutron number. 
Moreover, as clearly shown in Fig.~\ref{fig: Theor + Exp syst}(c), RQRPA and RQTBA yield on average more strength below the neutron threshold in comparison to the experimental data, demonstrating steadily decreasing TRK values toward $^{124}$Sn in the even-even isotopes.

The agreement of RQTBA  to experimental data, although improved compared to RQRPA, is still imperfect. 
This indicates that some mechanisms of the strength formation are still missing to achieve spectroscopic accuracy.  
A complete response theory should take into account the continuum, including the multiparticle escape, a more complete set of phonons (in particular, those of unnatural parity and isospin-flip), complex ground state correlations, and in principle higher-complexity configurations.  

\begin{figure*}[t]
  \subfigure{\label{subfig 1: Cs}}
  \subfigure{\label{subfig 2: Cs}}
  \subfigure{\label{subfig 3: Cs}}
  \subfigure{\label{subfig 4: Cs}}
  \subfigure{\label{subfig 5: Cs}}
  \subfigure{\label{subfig 6: Cs}}
  \subfigure{\label{subfig 7: Cs}}
  \subfigure{\label{subfig 8: Cs}}
  \subfigure{\label{subfig 9: Cs}}
\includegraphics[width=1.0\linewidth]{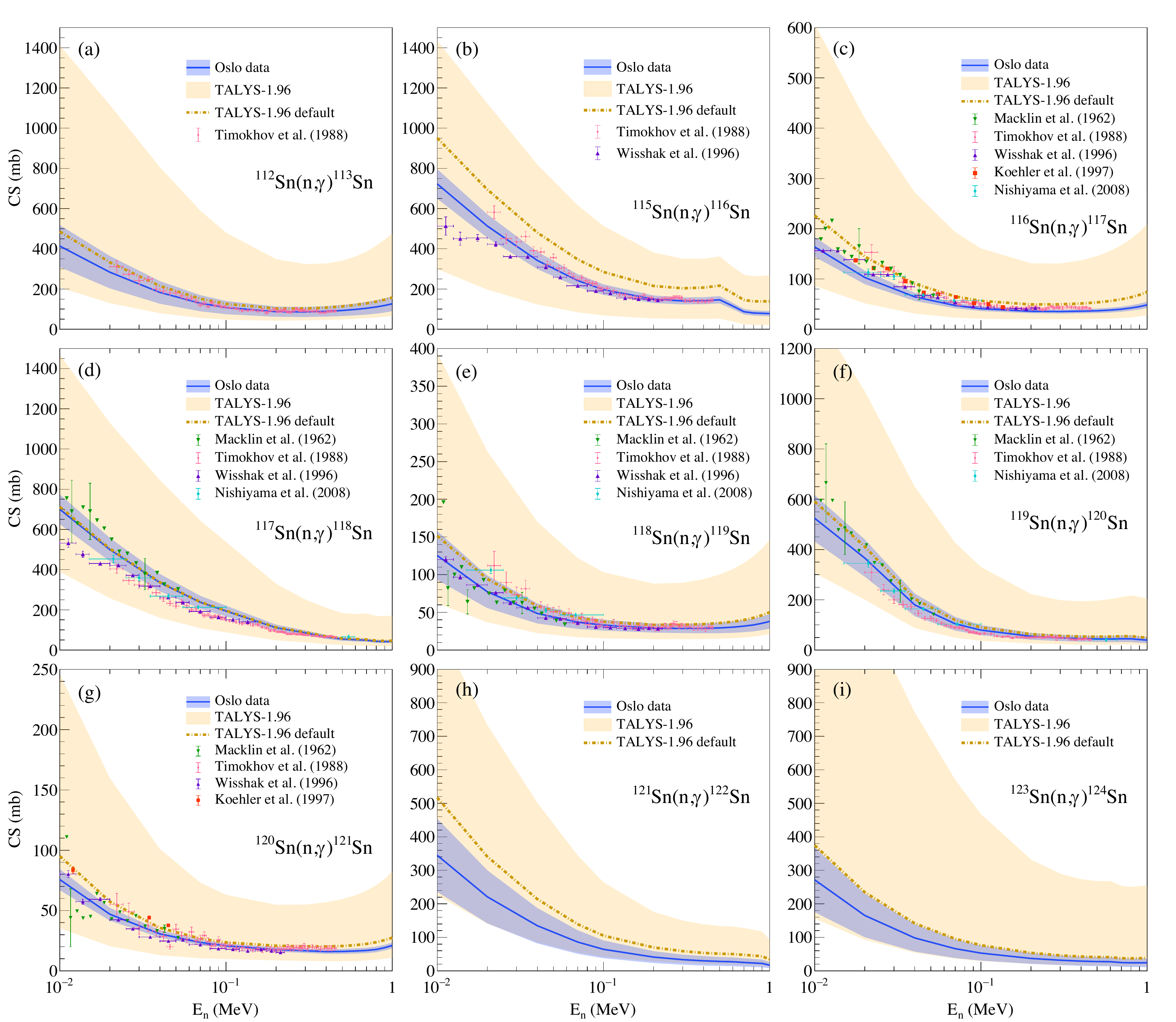}
\caption{\label{fig: CSs}
Cross sections (CS) for the $^{112}$Sn($n,\gamma)$$^{113}$Sn (a), $^{115}$Sn($n,\gamma)$$^{116}$Sn (b), $^{116}$Sn($n,\gamma)$$^{117}$Sn (c), $^{117}$Sn($n,\gamma)$$^{118}$Sn (d), $^{118}$Sn($n,\gamma)$$^{119}$Sn (e), $^{119}$Sn($n,\gamma)$$^{120}$Sn (f), $^{120}$Sn($n,\gamma)$$^{121}$Sn (g), $^{121}$Sn($n,\gamma)$$^{122}$Sn (h), and $^{123}$Sn($n,\gamma)$$^{124}$Sn (i) reactions. The predictions with the Oslo method inputs (blue bands) are compared with experimental data by Macklin \textit{et al.} \cite{Macklin1962}, Timokhov \textit{et al.} \cite{Timokhov1989}, Wisshak \textit{et al.} \cite{Wisshak1996}, Koehler \textit{et al.} \cite{Koehler2001}, Nishiyama \textit{et al.} \cite{Nishiyama2008}, and the TALYS uncertainty range obtained with different available GSFs, NLDs, and optical model potentials (beige band).
}
\end{figure*}
 
The single-particle continuum effect above the particle emission threshold can be taken into account by the smearing parameter, as it mainly causes uniform broadening of the individual peaks. 
This was quantified by direct calculations in Ref. \cite{Kamerdzhiev1998} that give a 100-200 keV width to characterize such a broadening, which is considerably smaller than the spreading width.
The two-fermionic cluster decomposition of the fully correlated dynamical kernel of the response function \cite{LitvinovaSchuck2019} suggests that the next-level complexity non-perturbative approximation is the $2q\otimes 2\rm phonon$ or correlated six-quasiparticle configurations in the intermediate propagators. 
The implementation of such configurations is becoming gradually possible with the increasing computational capabilities \cite{LitvinovaSchuck2019,Litvinova2023a}; however, systematic calculations for long isotopic chains of medium-heavy nuclei are still computationally demanding. 
Current efforts on optimizing the numerical $2q\otimes 2\rm phonon$ approach may enable such calculations in the near future.

\section{\label{sec 5: NCCS and MACS} Neutron capture cross sections}

\begin{figure*}[t]
\includegraphics[width=1.0\linewidth]{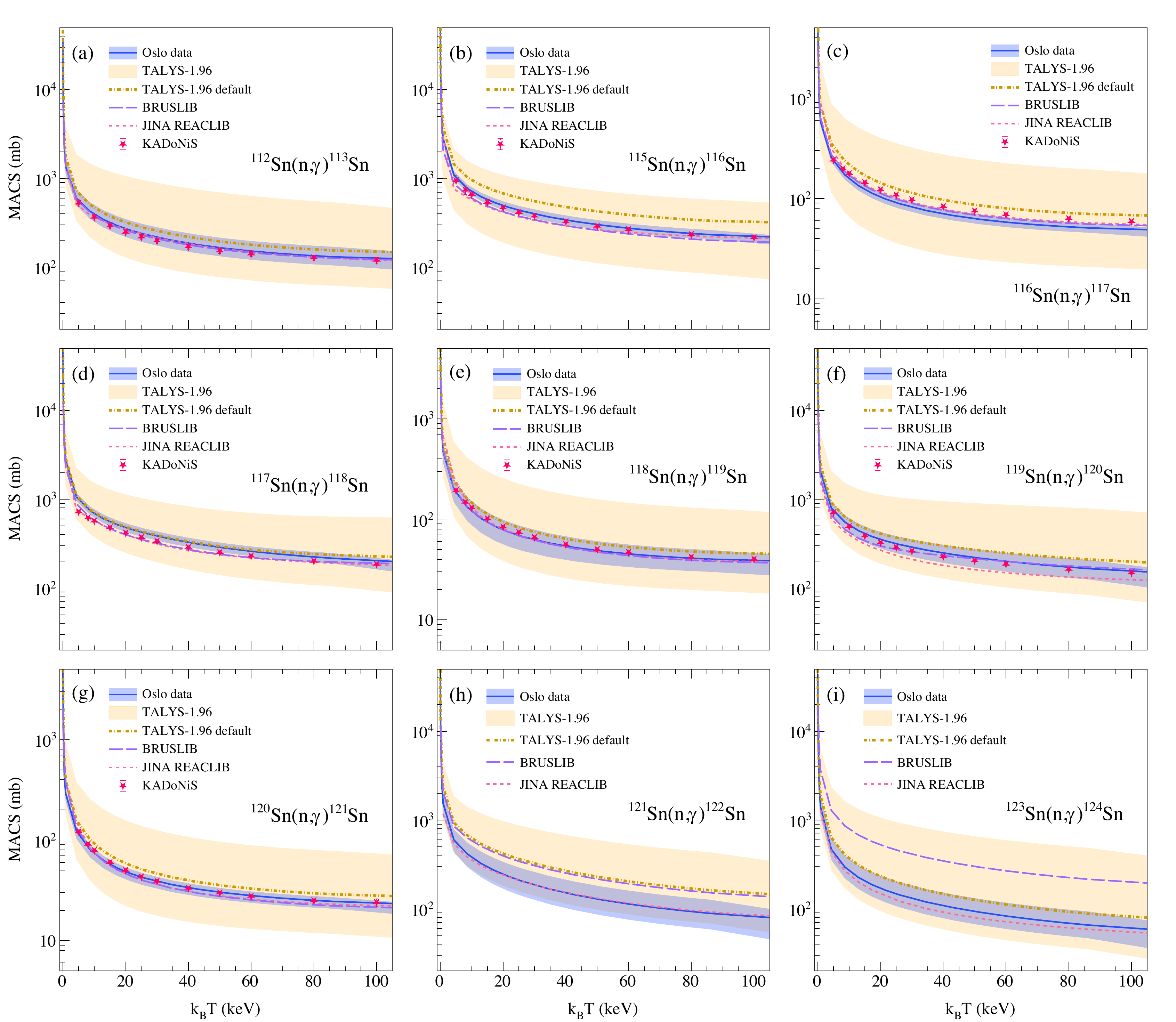}
\caption{\label{fig: MACSs}
Maxwellian-averaged cross sections (MACS) for the $^{112}$Sn($n,\gamma)$$^{113}$Sn (a), $^{115}$Sn($n,\gamma)$$^{116}$Sn (b), $^{116}$Sn($n,\gamma)$$^{117}$Sn (c), $^{117}$Sn($n,\gamma)$$^{118}$Sn (d), $^{118}$Sn($n,\gamma)$$^{119}$Sn (e), $^{119}$Sn($n,\gamma)$$^{120}$Sn (f), $^{120}$Sn($n,\gamma)$$^{121}$Sn (g), $^{121}$Sn($n,\gamma)$$^{122}$Sn (h), and $^{123}$Sn($n,\gamma)$$^{124}$Sn (i) reactions. The predictions with the Oslo method inputs (blue bands) are compared with recommended values from JINA REACLIB \cite{JINA}, BRUSLIB \cite{BRUSLIB}, KADoNiS \cite{KADONIS}, and the TALYS uncertainty range obtained with different available GSFs, NLDs, and optical model potentials (beige band).}
\end{figure*}

The experimental values of the NLD and GSF extracted with the Oslo method can further be used to estimate the radiative neutron-capture cross sections ($n,\gamma$) (NCCS) and reaction rates of interest for the astrophysical neutron capture processes. 
This was done within the statistical Hauser-Feshbach framework \cite{Hauser1952} with the TALYS reaction code (version 1.96) \cite{Koning2019,Koning2023}. 
The experimental Oslo method GSFs were combined with the $(p,p^{\prime})$ data at energies above their range to produce the tabulated $E1$ strengths used as  input functions. 
For the $M1$ input strength function, the $(p,p^{\prime})$ $M1$ data were chosen. 
For the optical model potential we use the phenomenological model of Koning and Delaroche \cite{Koning2003}. An alternative option provided by TALYS is the semimicroscopic Jeukenne-Lejeune-Mahaux model renormalized by the Bruy\`eres-le-Ch\^atel group \cite{Bauge2001}. In contrast to the earlier published cases of $^{165,166}$Ho \cite{Pogliano2023_2} and $^{185}$W \cite{Larsen2023}, the results obtained with both optical model potentials agree well within the uncertainty bands. Therefore, only calculations performed with the former model are presented in this work. The resulting NCCSs are presented in Fig.~\ref{fig: CSs}. The uncertainties due to the normalization parameters are included in the total uncertainty bands. The Oslo NCCSs (blue bands) are shown together with the span of TALYS cross sections, obtained by varying available GSFs, NLDs, and optical model potentials. In case of the Sn isotopes, since the radiative NCCS is rather insensitive to the optical potential and the experimental masses are adopted in each case, the former two are the major contributors to the wide spread of TALYS predictions, describing the overall discrepancies between NLD and GSF options available in TALYS \cite{Koning2023, Pogliano2023_2}. The cross section obtained with the default combination of models (constant temperature plus Fermi gas NLD model, SMLO form of the $E1$ strength, Koning and Delaroche global optical model potential) is also shown for each isotope in Fig.~\ref{fig: CSs}.

Even though the $^{112}$Sn($n,\gamma$)$^{113}$Sn reaction is of no potential interest for the astrophysical $s$ process, the comparison of the Oslo results with experimental ($n,\gamma$) cross sections from a comprehensive study by Timokhov \textit{et al.} \cite{Timokhov1989} is still valuable. Other experimental data in the keV region are also available for the neutron capture on the $^{115-120}$Sn targets [see Figs.~\ref{fig: CSs}(b)-(g)], covering almost all Sn isotopes involved in the $s$ process \cite{Koehler2001}. As was shown recently by Goriely \textit{et al} \cite{Goriely2021}, the ($n,\gamma$) reactions on $^{120,121,123}$Sn [Figs.~\ref{fig: CSs}(g)-(i)] might be of interest for the intermediate neutron capture process (i process). All experimental cross sections in Fig.~\ref{fig: CSs} were obtained with the time-of-flight method with neutrons produced in the $^7$Li($p,n$)$^7$Be reaction. The cross sections of Timokhov \textit{et al.} appear to be in excellent agreement with the Oslo cross sections for the ($n,\gamma$) reaction on $^{112}$Sn, $^{115}$Sn, $^{118}$Sn, $^{119}$Sn, and $^{120}$Sn. The Oslo NCCS is systematically lower for the $^{116}$Sn and systematically higher for the $^{117}$Sn targets compared to the data by Timokhov \textit{et al.}, while still agreeing within the uncertainty bands with the cross sections by Wisshak \textit{et al.} \cite{Wisshak1996}. In particular, a good agreement is achieved with the cross sections by Nishiyama \textit{et al.} for the neutron capture on $^{116-119}$Sn \cite{Nishiyama2008}. Overall, the Oslo results tend to agree within the uncertainty margins with all other experimental NCCS above neutron energies of $\approx 20-30$ keV. At lower energies, the experimental uncertainties increase, and different data sets demonstrate a wide spread of cross sections (of the order of $\approx 100$ mb). For the $^{121,123}$Sn targets no experimental data are available and, similarly to the lighter isotopes, the Oslo results are closer to the bottom part of the range of TALYS cross sections.

With the radiative NCCS at hand, the corresponding Maxwellian-averaged cross sections (MACS) can be estimated. The MACS values for the same target nuclei obtained with the experimental Oslo data are shown in Fig.~\ref{fig: MACSs} together with the span covered by available combinations of TALYS input models and the TALYS default MACSs. We also compare our results to the cross sections from the JINA REACLIB \cite{JINA} and BRUSLIB \cite{BRUSLIB} libraries, commonly used for astrophysical network calculations. The available data points from the KADoNiS database \cite{KADONIS} are also shown in Fig.~\ref{fig: MACSs}. It is important to note that all cross sections shown in Fig.~\ref{fig: MACSs} are stellar MACSs. For the $^{119}$Sn and $^{121}$Sn target nuclei the discrepancy between the stellar and laboratory MACSs might reach up to $\approx 56$\% and 20\%, respectively, below the thermal energy of 100 keV \cite{KADONIS}.

In the majority of considered cases, the recommended MACSs from the libraries fall well within the uncertainty bands of the Oslo results. The BRUSLIB MACSs for the $^{121}$Sn and $^{123}$Sn targets appear to be on average $\approx 1.2$ and $\approx 2.2$ times higher, respectively, compared to the Oslo MACSs for the thermal energies between 10 and 100 keV. This disagreement stems primarily from the combinations of the NLD model (Skyrme-Hartree-Fock-Bogoluybov plus Combinatorial NLDs \cite{Goriely2008}) and GSF model (Gogny-Hartree-Fock-Bogoliubov plus QRPA GSF \cite{Goriely2004}) employed in BRUSLIB. The latter model tends to underestimate the $E1$ strength distribution in the immediate vicinity of $S_n$. In general, the Skyrme-HFB plus combinatorial model reproduces the Oslo method NLDs quite well for the lighter Sn nuclei (e.g. $^{116}$Sn) and begins to overestimate the Oslo NLD values quite significantly toward more neutron-rich isotopes. Despite the model GSF being, on average, lower than the experimental strength, the net effect of these models combined together leads to the disagreement visible in Figs.~\ref{fig: MACSs}(h) and (i).

The Oslo MACSs agree quite well within the estimated uncertainties with the values provided by the KADoNiS database. Some systematic deviations are observed for the $^{116}$Sn and $^{117}$Sn targets, similarly to those in the NCCSs. The source of these deviations in both cases is not immediately obvious based on the used Oslo input data. The recommended KADoNiS values at 30 keV for Sn isotopes are largely based on the cross sections of Macklin \textit{et al.} \cite{Macklin1962}, Timokhov \textit{et al.}, Wisshak \textit{et al.}, Koehler \textit{et al.} \cite{Koehler2001}, Nishiyama \textit{et al.}, and are often presented at other $k_B T$ values by the average of evaluations from the ENDF/B-VII.1 \cite{Chadwick2011} and JENDL-4.0 \cite{JENDL} libraries. The uncertainties of these values, estimated for the majority of Sn isotopes from the deviations of these two evaluations, might be underestimating the KADoNiS systematic errors. The error bands of the Oslo MACSs, including all uncertainties due to the normalization of the nuclear inputs (NLD and GSF), provide far more conservative spans of the cross sections, nevertheless, considerably constraining the TALYS uncertainty range. 

\begin{figure}[t]
\includegraphics[width=1.0\linewidth]{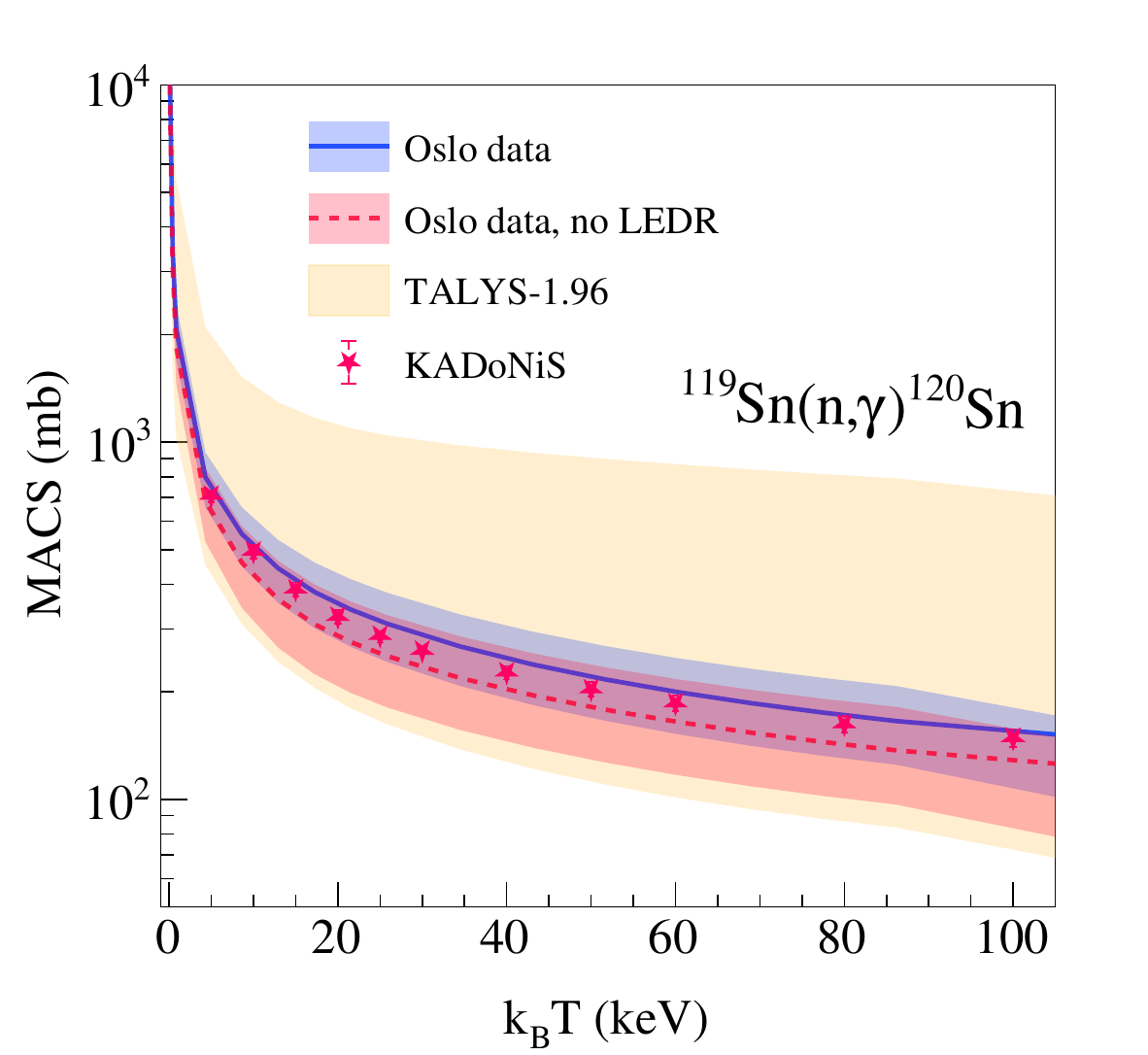}
\caption{\label{fig: MACSs no pdr}
MACS for the $^{119}$Sn($n,\gamma)$$^{120}$Sn reaction with and without the low-lying electric dipole strength included. The Oslo results with the LEDR (blue band) and without (red band) are compared with the TALYS uncertainty range (beige band) and recommended values from KADoNiS \cite{KADONIS}.
}
\end{figure}

As mentioned earlier, correct theoretical reproduction of the LEDR is of importance for astrophysical applications, in particular involving neutron-rich nuclei. Albeit the LEDR in the studied stable isotopes exhausts only $\approx 2-3\%$ of the TRK sum rule, it might still noticeably contribute to the radiative neutron capture rates and cross sections. To estimate the contribution of the observed LEDR in the Sn isotopes to the MACS, we performed TALYS calculations for the $^{119}$Sn($n,\gamma)$$^{120}$Sn reaction with the Oslo input GSF of $^{120}$Sn with an artificially subtracted LEDR, extracted according to the procedure in Sec.~\ref{subsec 3.2: GSFs and PDRs} (see Table \ref{tab:table_4} for the Gaussian peak parameters). This MACS is compared to the MACS extracted with the original Oslo GSF of $^{120}$Sn in Fig.~\ref{fig: MACSs no pdr}. The MACS obtained with no LEDR is consistently lower and, on average, it amounts to $\approx 80\%$ of the full MACS in the vicinity of 30 keV. Even though the cross sections overlap within the estimated uncertainty bands, the 20\% decrease is considerable for the relatively small exhausted fraction of the TRK sum rule ($\approx 3\%$). With the current status of available theoretical frameworks a consistent quantitative assessment of the role of the LEDR in astrophysical simulations remains a complex, non-trivial task, encouraging further advances in theoretical approaches and experimental studies of nuclei beyond the valley of stability.

\begin{table}[b]
\caption{\label{tab:astro_cases} Cases considered for the rates of the $^{121}$Sn($n,\gamma$)$^{122}$Sn and $^{123}$Sn($n,\gamma$)$^{124}$Sn reactions for the multi-zone stellar calculations.
}
\begin{ruledtabular}
\begin{tabular}{l|c|c}
 & $^{121}$Sn($n,\gamma$)$^{122}$Sn & $^{123}$Sn($n,\gamma$)$^{124}$Sn \\
\hline
\multicolumn{3}{c}{TALYS parameter uncertainties} \\
\hline
case 1 & min & min \\
case 2 & max & max \\
case 3 & min & max \\
case 4 & max & min \\
\hline
\multicolumn{3}{c}{TALYS model uncertainties} \\
\hline
case 5 & min & $-$ \\
case 6 & max& $-$ \\
case 7 & $-$ & min \\
case 8 & $-$ & max \\
\hline
\multicolumn{3}{c}{Experiment (this work)} \\
\hline
case 9 & min & min \\
case 10 & max & max \\
case 11 & min & max \\
case 12 & max & min \\
\end{tabular}
\end{ruledtabular}
\end{table}

\section{\label{sec 6: i process}Astrophysical implications}
To illustrate the impact of the newly determined reaction rates on some astrophysical applications, we consider the $i$-process nucleosynthesis in Asymptotic Giant Branch (AGB) stars. The AGB phase corresponds to the last evolutionary stage of $\sim 1 - 8$~$M_{\odot}$ stars \citep[e.g.][]{karakas14}. During this stage, hydrogen can be engulfed by one of the recurrent convective thermal pulses, leading to a proton ingestion event (PIE, e.g. \cite{fujimoto00,iwamoto04,suda10,stancliffe11}). During a PIE, protons are transported downwards in a timescale of about 1~hr and quickly burnt by the $^{12}$C($p,\gamma$)$^{13}$N reaction. The $^{13}$N isotope decays to $^{13}$C in a timescale of about 10~min. Then, the reaction $^{13}$C($\alpha,n$)$^{16}$O is activated, mostly at the bottom of the pulse, where $T \simeq 250$~MK. The neutron density goes up to about $10^{15}$~cm$^{-3}$ which leads to an $i$-process nucleosynthesis \citep[e.g.][]{cristallo09b,cristallo16,choplin21,choplin22a,choplin22b}. The $i$-process material is later dredged up to the stellar surface and expelled through stellar winds.

\begin{figure}[t]
\includegraphics[width=1.0\linewidth]{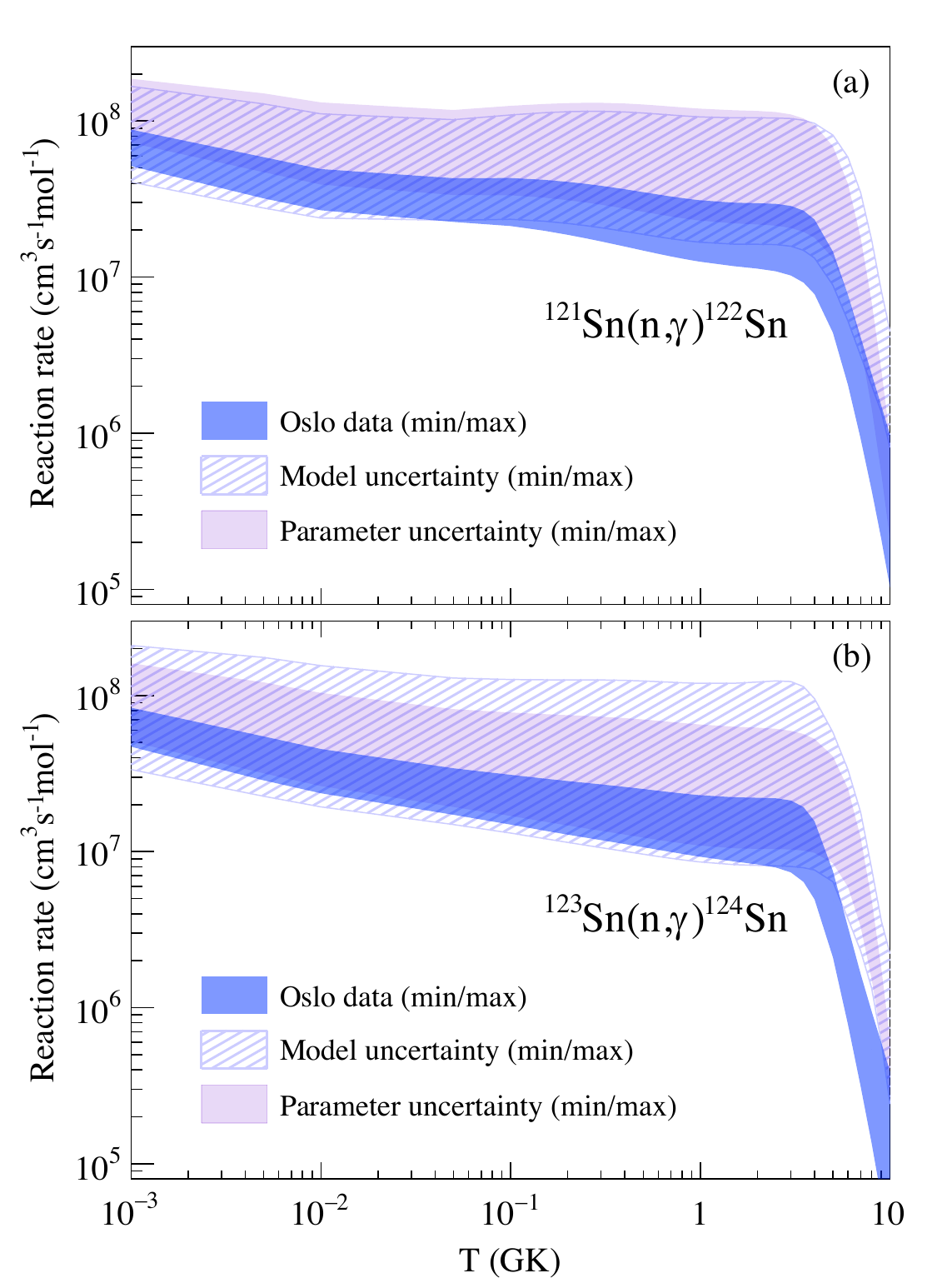}
\caption{\label{fig: Rates}
Uncertainty bands for the neutron capture rates in the $^{121}$Sn$(n,\gamma)$$^{122}$Sn (a) and $^{123}$Sn$(n,\gamma)$$^{124}$Sn reaction (b) reactions. The blue band corresponds to the span of experimentally constrained reaction rates due to uncertainties of the input Oslo NLD and GSF. The hatched band denotes the span of TALYS rates for all available GSF, NLD, and optical model potential combinations (model uncertainty). The purple band is due to the variation of the HFB+Combinatorial NLD and D1M+QRPA GSF model parameters according to the procedure of Ref.~\cite{martinet23} (parameter uncertainty). 
}
\end{figure}

Here, we investigated the impact of our new experimentally constrained $^{121,123}$Sn$(n,\gamma)$$^{122,124}$Sn reaction rates and corresponding uncertainties on the $i$-process nucleosynthesis in a 1~$M_{\odot}$ low-metallicity ([Fe/H]~$=-2.5$) AGB model computed with the STAREVOL code \cite{siess00,siess06,Goriely2021}. 
The network considered comprises 1160 nuclei, linked through 2123 nuclear reactions ($n$-, $p$-, $\alpha$-captures and $\alpha$-decays) and weak interactions (electron captures, $\beta$-decays).
The rates were extracted from the BRUSLIB database, the Nuclear Astrophysics Library of the Universit\'e Libre de Bruxelles\footnote{Available at http://www.astro.ulb.ac.be/bruslib/}\cite{arnould06} and the updated experimental and theoretical rates from the NETGEN interface \cite{BRUSLIB}. 
Additional information on the stellar physics ingredients, modeling, and nuclear physics can be found in Refs.~\cite{choplin21,Goriely2021,choplin22a}. 

\begin{figure}[t]
\includegraphics[width=1.0\linewidth]{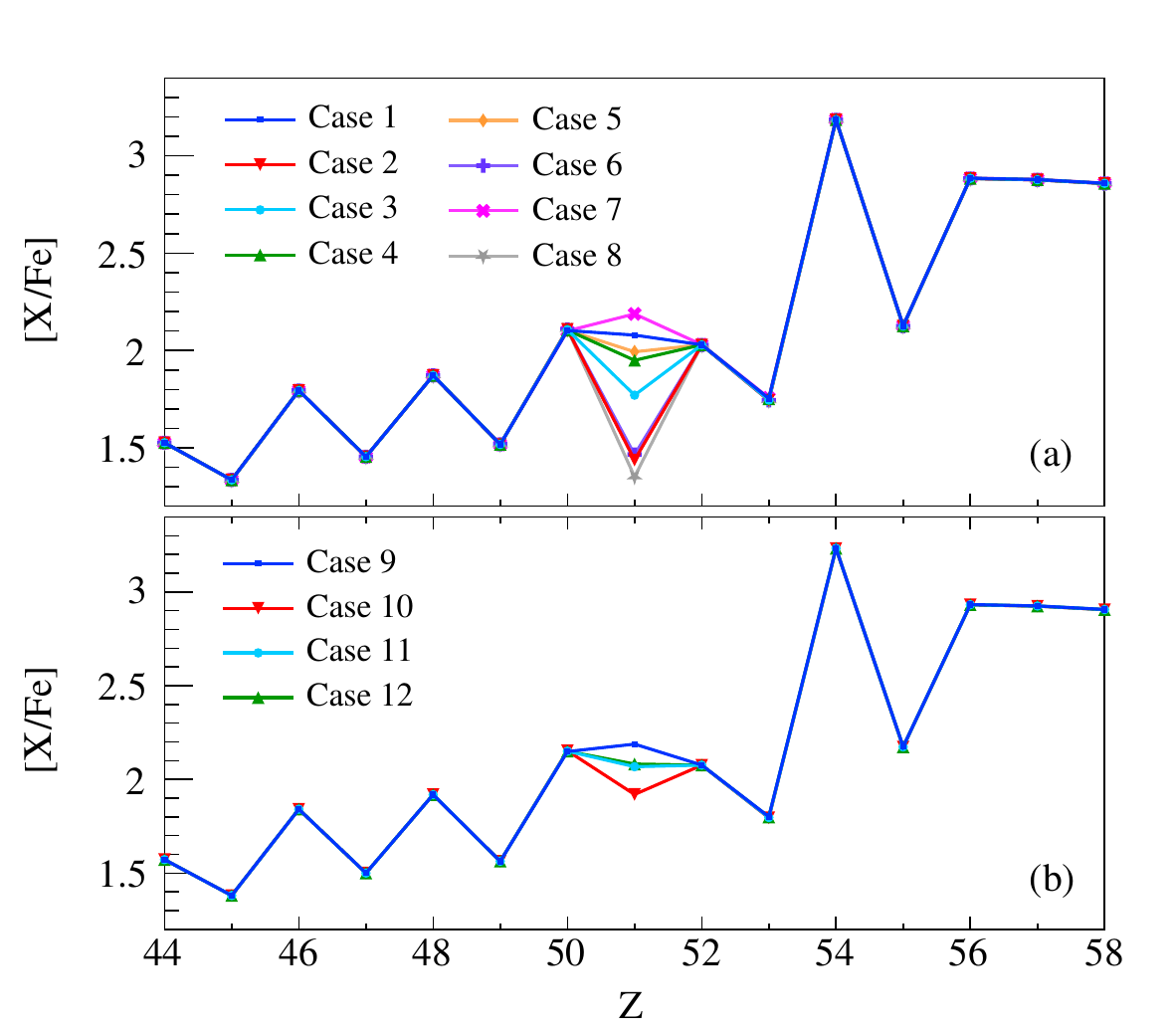}
\caption{\label{fig: Abundances elements}
Final surface elemental abundances (after decays) of multi-zone AGB stellar models experiencing $i$-process nucleosynthesis, computed with different combinations of $^{121,123}$Sn$(n,\gamma)$ rates. Shown are the [X/Fe] ratios defined as [X/Fe] = $\log_{10}(N_{\rm X}/N_{\rm Fe})_{\star} - \log_{10}(N_{\rm X}/N_{\rm Fe})_{\odot}$ with $N_{\rm X}$ the number density of an element X. The first and second $\log_{10}$ terms refer to the abundances of the model and the Sun, respectively.
(a) Eight theoretical rates combinations are considered. These are theoretical estimates of parameter and model uncertainties affecting TALYS predictions of $^{121}$Sn$(n,\gamma)$ and $^{123}$Sn$(n,\gamma)$ reaction rates. (b) Same but with the new experimentally constrained rates from this work. 
}
\end{figure}

\begin{figure}[t]
\includegraphics[width=1.0\linewidth]{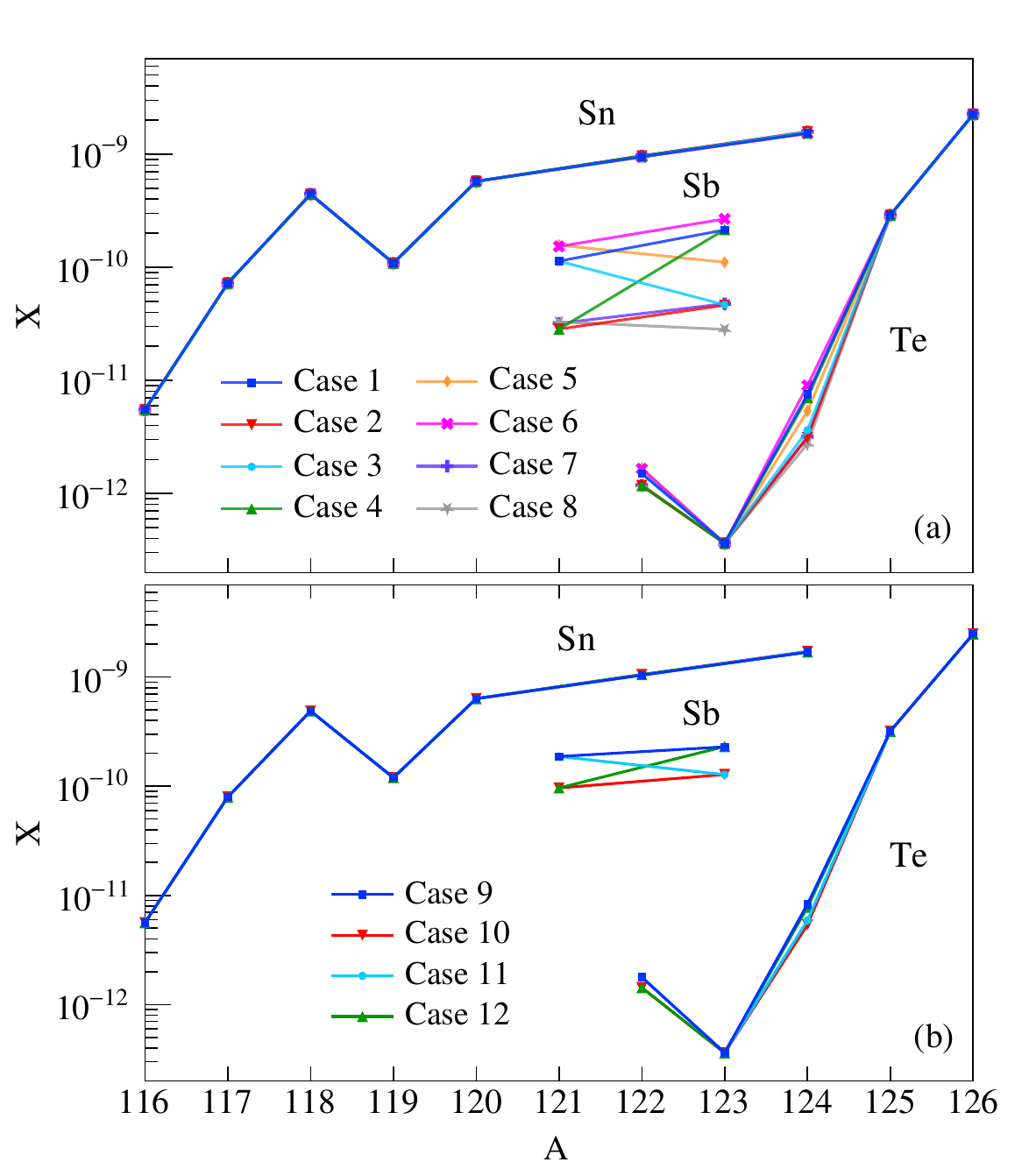}
\caption{\label{fig: Abundances isotops}
Same as Fig.~\ref{fig: Abundances elements} but for the isotopic mass fraction $X$ as a function of the mass number $A$, around the Sn region. 
}
\end{figure}

Since relatively accurate MACS have been previously measured for stable Sn isotopes, we only consider here the uncertainties affecting $^{121}$Sn$(n,\gamma)$ and $^{123}$Sn$(n,\gamma)$ reaction rates. To do so, we first considered their theoretical TALYS predictions and associated uncertainties. The latter include both parameter and models uncertainties, as extensively discussed in Ref.~\cite{martinet23}. The impact of uncorrelated parameter uncertainties has been estimated considering four cases (cases 1--4 in Table~\ref{tab:astro_cases}) obtained on the basis of the HFB+Combinatorial NLD and D1M+QRPA GSF models (note that similar parameter uncertainties are obtained for different NLD or GSF models, as discussed in Ref.~\cite{martinet23}). The four cases correspond to the different minimum / maximum possible combinations for both rates.
The impact of correlated nuclear model uncertainties was investigated by considering the NLD and GSF models leading to the lower or upper limits of the TALYS MACSs [beige bands in Fig.~\ref{fig: MACSs}(h) and (i)] and, thus, the respective reaction rates; these correspond to cases 5 to 8 in Table~\ref{tab:astro_cases}. 
Finally, the new experimentally constrained rates are considered (cases 9 to 12) and assumed to be uncorrelated. In total, 12 AGB simulations experiencing a PIE were computed with these different rate combinations. The associated uncertainty bands are shown in Fig.~\ref{fig: Rates}.

As seen in Fig.~\ref{fig: Abundances elements}, the impact of $^{121,123}$Sn$(n,\gamma)$ rate uncertainties on the $i$-process nucleosynthesis is local, arising at $Z=51$ (Sb). 
This is in line with recent results from \cite{martinet23} where the relevant reactions for $i$-process nucleosynthesis in AGB stars were shown to mainly have a local impact on the resulting chemical abundances. In particular, the $^{121,123}$Sn$(n,\gamma)$ reactions appear in their Table~1, listing the key uncertain reactions affecting $i$-process predictions. 
The impacted element, Sb ($Z=51$), is produced in different quantities depending on the strength of the $^{121,123}$Sn$(n,\gamma)$ reactions, the latter competing with $\beta$-decay (with half-lives of 27~hr and 129~days, respectively). 
A minimal rate for both $^{121,123}$Sn$(n,\gamma)$ reactions favors the production of $^{121,123}$Sb through $\beta$-decay (blue pattern in Fig.~\ref{fig: Abundances isotops}). By contrast, if the $(n,\gamma)$ rates are maximal, the flow favours the production of $^{122,124}$Sn and less of $^{121,123}$Sb, which results in lower $^{121,123}$Sb abundances (red pattern in Fig.~\ref{fig: Abundances isotops}).
Globally, TALYS theoretical uncertainties lead to an uncertainty of 0.84 dex in the final surface Sb abundance (the uncertainty is 0.65 dex for parameter uncertainties and 0.84 dex for model uncertainties). It is decreased to 0.27 dex when considering the new experimentally constrained rates (Fig.~\ref{fig: Abundances elements}). 
As shown in Fig.~\ref{fig: Abundances isotops}, for the $^{121}$Sb ($^{123}$Sb) isotope, the overall uncertainty is reduced from 0.75 (0.98) to 0.29 (0.26) dex.

\section{\label{sec 6: Conclusion}Conclusions and outlook}

In this work, a consistent analysis of the $^{111-113,116-122,124}$Sn isotopes with the Oslo method applied to light-particle-induced reaction data has been presented. 
The extracted NLDs demonstrate a clear constant-temperature trend below the neutron separation energy. 
They appear to be in good agreement with each other and reproduce the low-lying discrete states up to $\approx 2-3$ MeV quite well. 
The Oslo GSFs are fully compatible with the available Coulomb excitation $(p,p^{\prime})$ data for even-even isotopes $^{112,116,118,120,124}$Sn within the uncertainty bands and demonstrate a smooth evolution of the low-lying dipole strength with a slight increase of the GSF slope toward the heaviest studied $^{124}$Sn isotope. 

Based on the Oslo method and $(p,p^{\prime})$ strength distributions, the low-lying $E1$ strength on top of the IVGDR was found to be located at $\approx 7.8-8.3$ MeV and exhaust $\approx 2-3\%$ of the TRK sum rule for all the studied nuclei.
The observed trend does not reveal any strong dependence on  neutron number, and suggests a local maximum of strength at $^{120}$Sn. 
The 6.4-MeV component of the LEDR extracted in $^{118-122,124}$Sn demonstrates an approximate dependence on neutron excess and might potentially be related to a similar concentration of strength observed in this energy region in earlier works \cite{Endres2012, Pellegri2014}, where it was interpreted as the PDR. 

The experimental results have been compared to calculations of the LEDR in the even-even Sn isotopes within the RQRPA and RQTBA frameworks. Despite a greatly improved agreement with the experimental strength distribution within the PDR and IVGDR regions as compared to the RQRPA approach, the RQTBA calculations do not reproduce the experimental TRK values extracted within the same energy regions. 
Both the RQRPA and RQTBA predict a clear linear increase in strength at $\approx 8-10$ MeV toward $^{124}$Sn, in contrast to the experimental estimates that are approximately constant throughout the whole chain of the investigated isotopes. 

The Oslo method NLDs and GSFs were further used to constrain the radiative neutron-capture cross sections and Maxwellian-averaged  cross sections with the reaction code TALYS. 
Overall, the obtained values are in good agreement with available experimental data and Maxwellian-averaged cross sections and reaction rates reported in the JINA REACLIB, BRUSLIB, and KADoNiS libraries. 
The $^{121,123}$Sn$(n,\gamma)$ reaction rates obtained with the Oslo input NLDs and GSFs were found to locally impact the production of $^{121,123}$Sb in the $i$-process nucleosynthesis in AGB stars and significantly reduce the available model and parameter uncertainties for the estimated final surface Sb abundance. 

Despite the relatively small fractions of the TRK sum rule exhausted in the studied stable Sn isotopes, the low-lying dipole strength in these nuclei has a noticeable impact on the estimated reaction cross sections and rates. 
Further improvements in the microscopic calculations of $E1$ and $M1$ strength distributions close to the neutron threshold, especially in neutron-rich nuclei beyond the valley of stability, are highly desirable for future astrophysical calculations involving the $i$- and $r$-process nucleosynthesis. 
Moreover, from a nuclear-structure point of view, further detailed studies of the underlying structure of the states contributing to the LEDR strength are called for. 
In the near future, high-resolution experiments utilizing ($d,p\gamma$) and ($p,d\gamma$) reactions to populate the same even-even nucleus $^{118}$Sn are envisaged to shed new light on this very intriguing issue. 

\begin{acknowledgments}
The authors express their thanks to J.~C.~M\"{u}ller, P.~A.~Sobas, V. Modamio and J.~C.~Wikne at the Oslo Cyclotron Laboratory for operating the cyclotron and providing excellent experimental conditions. 
The authors are also thankful to A. Görgen and G. M. Tveten for their contribution to the experiment and discussions. 
A.~C.~L. gratefully acknowledges funding by the European Research Council through ERC-STG-2014 under Grant Agreement No.\ 637686, and from the Research Council of Norway, project number 316116. S. S. acknowledges funding from  the Research Council of Norway, project numbers 325714 and 263030. 
P.v.N.-C. acknowledges support by the Deutsche Forschungsgemeinschaft (DFG, German Research Foundation) under Grant No. SFB 1245 (Project ID No. 279384907). The work of E.L. was supported by the GANIL Visitor Program, US-NSF Grant PHY-2209376, and US-NSF Career Grant PHY-1654379. S.M. and S.G. acknowledge support from the European Union (ChETEC-INFRA, project No. 101008324). This work was supported by the F.R.S.-FNRS under Grant No. IISN 4.4502.19. L.S. and S.G. are senior F.R.S.-FNRS research associates; A.~C. is F.R.S.-FNRS postdoctoral researcher.

\end{acknowledgments}


\bibliography{tin_2023}
\end{document}